\documentclass[lettersize,journal]{IEEEtran}
\usepackage{array}
\usepackage{stfloats}
\usepackage{url}
\usepackage{verbatim}
\usepackage{graphicx,subcaption}
\usepackage[T1]{fontenc}
\usepackage{amsmath,amssymb,amsfonts,dsfont,breqn}
\usepackage{graphicx}
\usepackage[dvipsnames]{xcolor}
\usepackage{algorithmic,booktabs}
\usepackage{textcomp}
\usepackage{upgreek}
\usepackage[ruled,vlined]{algorithm2e}
\usepackage{bm,soul}
\usepackage{multirow}

\newcommand{\HIDEFIG}[1]{}

\newtheorem{remark}{Remark}

\newtheorem{assumption}{Assumption}

\newtheorem{Proposition}{Proposition}

\def\BibTeX{{\rm B\kern-.05em{\sc i\kern-.025em b}\kern-.08em
    T\kern-.1667em\lower.7ex\hbox{E}\kern-.125emX}}
\def\BibTeX{{\rm B\kern-.05em{\sc i\kern-.025em b}\kern-.08em
    T\kern-.1667em\lower.7ex\hbox{E}\kern-.125emX}}
\usepackage{balance}
\begin{document}
\title{Dynamic Traffic Modeling using Equivalent Circuit Models}
\author{Joseph Moyalan, \IEEEmembership{Member, IEEE}, Ricardo de Castro, \IEEEmembership{Senior Member, IEEE}, Iman Ebrahimi, Xuchang Tang, Xinfan Lin, \IEEEmembership{Senior Member, IEEE}
\thanks{This work is supported by the California Climate Action Seed Grants of the University of California (Grant No. R02CP6996).
Joseph Moyalan, Iman Ebrahimi, and Ricardo de Castro are with the Department of Mechanical and Aerospace Engineering, UC Merced. Xuchang Tang and Xinfan Lin are with the Department of Mechanical and Aerospace Engineering, UC Davis.}}

\maketitle
\markboth{IEEE TRANSACTIONS ON INTELLIGENT TRANSPORTATION SYSTEMS}%
{Moyalan \MakeLowercase{\textit{et al.}}: ECM--Based EV Evacuation With Mobile Charging Stations}
\begin{abstract}
This paper presents an equivalent circuit model (ECM) for dynamic traffic modeling by exploiting analogies between electrical and transportation variables. Vehicle concentration is represented by capacitor voltage, while traffic flow is modeled using controllable current sources. The proposed formulation is shown to preserve the conservation and sending–receiving relationships of the spatially discretized Lighthill–Whitham–Richards traffic model. The framework is extended to restricted inflow, traffic signals, and two-road merging and diverging scenarios. Validation against microscopic Simulation of Urban MObility (SUMO) simulations, involving traffic signals, two-road merging, and single-road diverging scenarios, demonstrates that the proposed ECM successfully reproduces the principal macroscopic traffic dynamics, including queue formation, backward congestion propagation, vehicle accumulation, saturation, and subsequent traffic recovery.
\end{abstract}

\section{Introduction}
The increasing complexity of transportation networks has motivated the development of modeling frameworks that can describe traffic dynamics while remaining sufficiently compact and interpretable for analysis, optimization, and control. Conventional traffic-flow models span a broad range of resolutions, from microscopic approaches that explicitly describe individual vehicle interactions to macroscopic models that characterize traffic through aggregate quantities such as vehicle density, flow, and average speed \cite{hamilton2013evolution,othman2019ecological,pasquale2019traffic}. Among macroscopic approaches, the Lighthill–Whitham–Richards (LWR) model provides a well-established description of traffic dynamics based on conservation laws and a fundamental relationship between traffic flow and vehicle density \cite{ferrara2018freeway}. Although such physics-based models provide an effective representation of congestion formation and propagation, their partial-differential-equation formulation and associated network coupling can complicate their direct integration with optimization and control methodologies, particularly for transportation networks involving multiple links, traffic-control devices, and interacting traffic streams.

Equivalent circuit models (ECMs) provide an alternative modeling paradigm in which the governing dynamics of a physical system are represented using analogous electrical variables and circuit components. Such representations are not limited to electrical systems. For example, electrical–thermal analogies have been extensively employed for thermal modeling, where temperature is represented by voltage, heat flow by electrical current, thermal resistance by electrical resistance, and thermal storage by capacitance \cite{wunsche1997electro}. These analogies enable distributed heat-transfer processes to be approximated by interconnected lumped circuit elements, providing computationally convenient models for both steady-state and transient thermal analysis. The resulting circuit representation also permits the use of mature tools from circuit theory, simulation, optimization, and control while retaining a direct physical interpretation of the underlying energy-storage and transport mechanisms. Similar analogical modeling approaches have therefore been adopted in several non-electrical domains whenever the underlying system is governed by conservation and transport relationships \cite{amanuel2021design,amir2022dynamic}.

The movement of vehicles through a transportation network exhibits several characteristics that naturally motivate an analogous circuit representation. Traffic flow describes the number of vehicles crossing a location per unit time and is therefore conceptually analogous to electrical current, which describes the transport of electric charge per unit time. Similarly, the accumulation of vehicles within a road segment can be interpreted as a storage phenomenon analogous to the storage of electric charge in a capacitor. These similarities have motivated previous investigations of electrical-circuit analogies for transportation networks. Early and subsequent studies employed Kirchhoff-type relationships and resistive or nonlinear circuit networks for traffic assignment and the analysis of congested transportation networks \cite{furber1973electrical,wellin1975simulation}. More recently, circuit-based representations have been investigated for roadway topology and route-flow analysis \cite{callejas2020exploring,sinop2023robust,moyalan2026optimal}. Such studies demonstrate that electrical analogies can provide useful mathematical and computational interpretations of transportation-network flow.

Nevertheless, much of the existing circuit-based traffic literature has primarily addressed static or steady-state traffic assignment, route selection \cite{li2019modelling,han2024optimal}, network topology \cite{tolmachev2019transport}, or equilibrium flow relationships \cite{danchuk2019simulation}.  In these formulations, circuit quantities are generally used to reproduce steady-state network flow conservation or link-cost relationships rather than dynamic temporal evolution of vehicle accumulation and congestion within individual road segments. Consequently, a systematic dynamic ECM formulation remains comparatively underexplored. Establishing such a correspondence is particularly important if circuit-based traffic models are to capture dynamic phenomena such as queue formation and backward propagation of congestion.

To address this gap, this paper proposes an ECM for dynamic macroscopic traffic modeling by establishing explicit analogies between electrical and transportation variables. We spatially discretize each road link into cells, representing the total number of vehicles in each cell as the voltage across a unit-capacitance capacitor. Traffic flow between adjacent cells is represented using controllable current sources whose values depend nonlinearly on the vehicle concentrations of the upstream and downstream cells. The current-source characteristics are constructed according to the sending and receiving constraints associated with a trapezoidal traffic fundamental diagram  \cite{boyles2020transportation}. As a result, the capacitor current-balance equation reproduces vehicle conservation, while the nonlinear current-source law reproduces the flow constraints of the spatially discretized LWR traffic model. This establishes a direct physical and mathematical correspondence between the proposed circuit representation and conventional first-order macroscopic traffic dynamics.


The main contributions of this paper are summarized as follows:

1. {\bf Dynamic traffic ECM and connection with the LWR model:} A nonlinear RC-circuit representation of dynamic traffic flow is developed and shown to be mathematically equivalent to the spatially discretized LWR model with a trapezoidal fundamental diagram, preserving vehicle conservation and traffic flow constraints.

2. {\bf Queue and network flow modeling:} The ECM is extended to model restricted inflow, traffic signals, and merging and diverging scenarios, incorporating external queue dynamics and optimization-based flow allocation to enforce downstream receiving constraints and accommodate desired flow priorities.

3. {\bf Microscopic validation:} The proposed ECM is evaluated against microscopic traffic simulations generated using Simulation of Urban MObility (SUMO). The comparison investigates traffic-signal, two-road merge, and single-lane diverge scenarios and demonstrates the capability of the ECM to reproduce the principal macroscopic traffic phenomena, including queue formation, upstream congestion propagation, saturation, and traffic recovery.

The remainder of the paper is organized as follows. Section~\ref{section:prelim} presents the preliminaries, notation, and problem formulation. The main results, including the ECM-based formulation of the evacuation problem, are detailed in Section~\ref{section:main}. Validation results with SUMO are provided in Section~\ref{section:results}, followed by concluding remarks.
\section{Preliminaries}\label{section:prelim}
This section introduces the basic concepts of Dynamic Traffic Assignment (DTA), including link and node models.
\subsection{Transportation Network Laws}\label{section:transport_laws}
\begin{figure}[ht]
    \centering
    \includegraphics[width=0.75\linewidth]{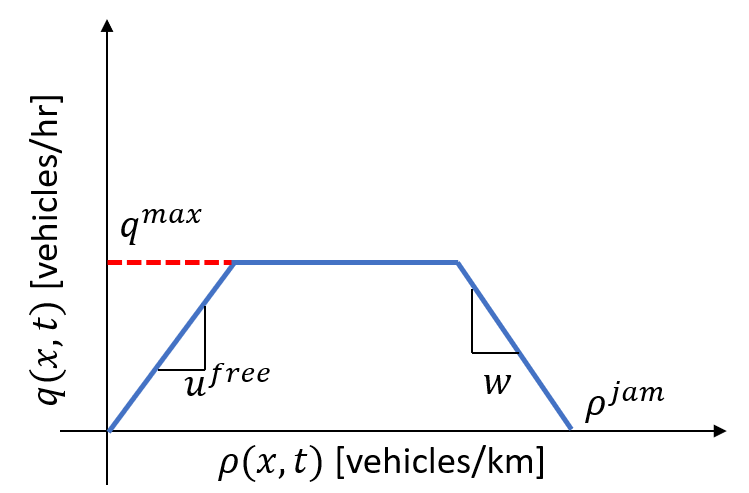}
    \caption{Illustration of the fundamental macroscopic relationship in a traffic link, assuming a trapezoidal shape. }
    \label{fig:funda_diag}
\end{figure}

The fundamental (macroscopic) relationship \cite{ferrara2018freeway}  in a traffic link is given by
\begin{align}
    q(x,t) = \rho(x,t) u(x,t) \label{eq:dta_fund_eq}
\end{align}
where $q, \rho$, and $u$ are continuous piecewise differentiable functions of position ($x$) and time ($t$), and represent traffic flow [vehicle/hr], traffic density [vehicle/km], and average speed [km/hr] in a traffic link, respectively. The second fundamental relationship is the continuity or conservation equation \cite{ferrara2018freeway} and is given by
\begin{align}
    \frac{\partial\rho}{\partial t}+\frac{\partial q}{\partial x}=0. \label{eq:dta_conti_eq}
\end{align}
The third and final relationship is derived from the Lighthill–Whitham–Richards (LWR) model \cite{ferrara2018freeway}. The LWR model formulates a direct relation between $q$ and $\rho$ based on the Fundamental Diagram of a traffic link in steady-state conditions. This relation translates to
\begin{align}
    q(x,t) = f(\rho(x,t)) \label{eq:q_rho_relation}
\end{align}
for some function $f(\rho)$. In this paper, we assume that $f$ takes a trapezoidal shape as shown in Fig.~\ref{fig:funda_diag}  \cite{ferrara2018freeway}. The equation of $f(\rho)$ is given as follows:
\begin{align}
    f(\rho) = \min\{ u^{free}\rho,q^{max},w(\rho^{jam}-\rho)\}. \label{eq:fund_diag}
\end{align}
Here, $q^{max}$ represents the maximum vehicle flow that the link can support without congestion. We denote the slope of the first and second half of the diagram by $u^{free}$ (free flow speed on the link) and $w$ (backward wave speed), respectively. Also,  $\rho^{jam}$ denotes the jamming vehicle density on the link.

\subsection{Link and Node Model}
\begin{figure}[ht]
    \centering
    \includegraphics[width=0.97\linewidth]{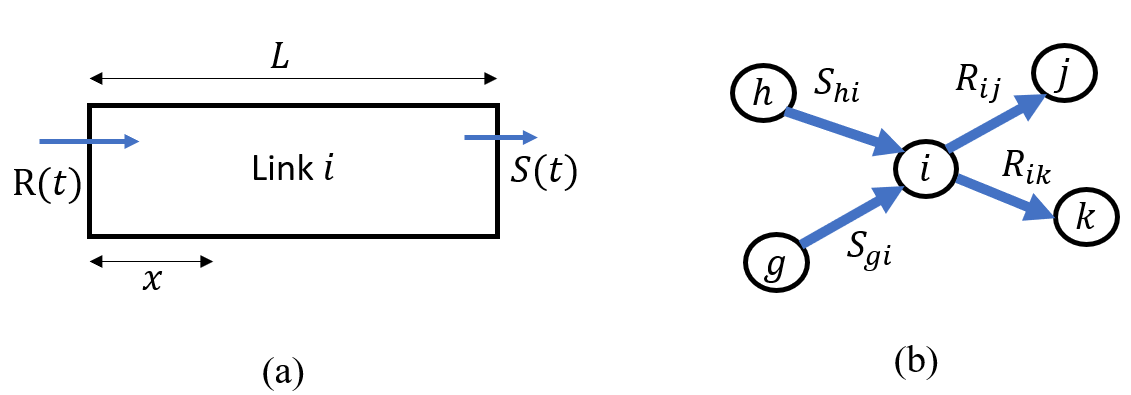}
    \caption{Illustration of (a) link and (b) node model. }
    \label{fig:link_node_model}
\end{figure}
The length of a single link is denoted by $L$. Therefore, $x=0$ and $x=L$ represent the upstream end and downstream end of the link, respectively (Fig.~\ref{fig:link_node_model}a). The input vehicle flow to the link is denoted by $R(t)$. Similarly, $S(t)$ denotes the output vehicle flow.

The node model complements the link model discussed in this paper by capturing the interactions of flows from multiple incoming links. The node model associated with node \( i \) takes as inputs the sending flows \( S_{hi}(t) \) for each incoming link \( (h,i) \in \Gamma^{-1}(i) \) and gives output as the receiving flows \( R_{ij}(t) \) for each outgoing link \( (i,j) \in \Gamma(i) \) (Fig.~\ref{fig:link_node_model}b). Therefore,
\begin{align*}
    \sum_{(h,i)\in \Gamma^{-1}(i)} S_{hi}(t) = \sum_{(i,j)\in \Gamma(i)}R_{ij}(t)
\end{align*}
\subsection{Discretizations over Space}
The transportation network laws described in Section~\ref{section:transport_laws} are continuous in both space and time. However, in this paper, the spatial domain is discretized by dividing each road link into multiple cells, allowing the vehicle concentration to be evaluated at different locations along the road. The length of each cell is defined by the spatial interval $\Delta x$. This spatial discretization enables the density function $\rho(x,t)$ to be discretized as follows:
\begin{align}
n_i(t) = \int_{x_{i-0.5}}^{x_{i+0.5}}\rho(x,t)dx \label{eq:approx}
\end{align} 
where $n_i(t)$ represents the total number of vehicles present within the $i^{th}$ cell at time $t$. Here, $x_{i-0.5}$ and $x_{i+0.5}$ represent the left/upstream boundary and right/downstream boundary of cell $i$, respectively.
\begin{assumption}\label{assume:const_density}
    We assume that vehicle density inside a cell is uniform such that
    \begin{align}
        \rho_i(t) \approx \frac{n_i(t)}{\Delta x} 
    \end{align}
\end{assumption}
\section{Dynamic Equivalent Circuit for Traffic Model}\label{section:main}
In this section, we propose an equivalent circuit model (ECM) to represent dynamic traffic behavior.
\subsection{Analogy of Circuit and Traffic Variables}
We start by revising the definitions of electrical variables and components and highlighting the natural similarities with traffic-domain variables. We use the notation  $x \leftrightarrow y$ to indicate that the (circuit) variable $x$ is an $``$analog" of the (traffic) variable $y$. 

Recall that electric current $I =  dQ / dt$ represents the amount of charge $ Q$ that crosses a cross-sectional area during unit time $dt$. This definition is analogous to traffic flow $q = dn / d t$, where $ n$ is the number of vehicles that pass through a fixed point over $d t$. Next, we revisit the definition of a capacitor, an electric device capable of storing electric charge. 
Physically, the amount of charge that the capacitor stores ($Q$) is proportional to the voltage difference ($V$) between its terminals, i.e., $Q = CV$, where $C$ is the constant of proportionality known as capacitance. Therefore, we conclude that $I \leftrightarrow q$ and $Q \leftrightarrow CV \leftrightarrow n$. In the proposed ECM framework, the capacitance of each road cell can be defined using two alternative approaches, depending on the desired interpretation of the capacitor voltage:

1. {\bf Unitary capacitance:} By assigning a unitary capacitance ($C=1$) to each road cell, the capacitor relationship $ Q=C V$ reduces to $ V= Q$. Under this representation, the capacitor voltage is analogous to the total number of vehicles stored within the corresponding road cell. Consequently, the voltage $V_i(t)$ represents the vehicle concentration $n_i(t)$, expressed as the number of vehicles present in cell $i$ at time $t$.

2. {\bf Distance-based capacitance:} Alternatively, the capacitance can be defined proportional to the length of the corresponding road cell, i.e., $C=\Delta x$, where $\Delta x$ denotes the cell length. The capacitor relationship then becomes $ V= Q/\Delta x$. Under this representation, the capacitor voltage is analogous to the vehicle density $\rho_i(t)=n_i(t)/\Delta x$, expressed in vehicles per unit distance. Thus, the capacitance accounts for the spatial extent of the road cell, while its voltage directly represents the corresponding traffic density.

Both approaches provide consistent representations of vehicle accumulation and conservation within the ECM framework. The key difference lies in the physical interpretation of the capacitor voltage: the unitary-capacitance approach represents the total number of vehicles in a cell, whereas the distance-based approach represents vehicle density. In this work, we adopt the unitary-capacitance representation ($C=1$), such that the capacitor voltage directly corresponds to the number of vehicles stored within each road cell. This choice simplifies the circuit formulation while preserving the underlying traffic conservation relationships.
\subsection{Proposed ECM}\label{section:ECM_proposed}
\begin{figure}[ht]
    \centering
    \includegraphics[width=0.97\linewidth]{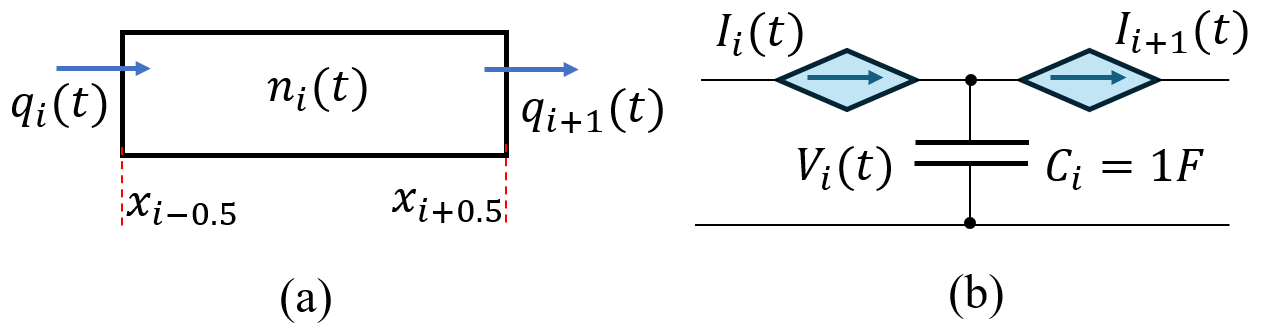}
    \caption{Illustration of (a) a single cell and (b) the corresponding ECM. }
    \label{fig:concentration_diagram}
\end{figure}

\begin{figure}[ht]
    \centering
    \includegraphics[width=0.97\linewidth]{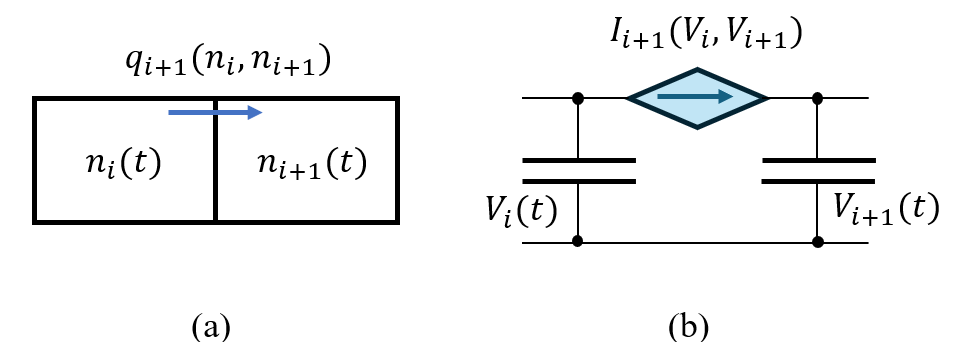}
    \caption{Illustration of (a) flow between two cells and (b) the corresponding ECM. }
    \label{fig:flow_diagram}
\end{figure}

Fig.~\ref{fig:concentration_diagram} and \ref{fig:flow_diagram} together represent the proposed ECM for the traffic dynamics where each road cell is of unit length. In Fig.~\ref{fig:concentration_diagram}, we observe that the concentration of a road cell ($n_i(t)$) is represented by the voltage difference ($V_i(t)$) across the capacitor $C_i$ in the ECM. Similarly, the input ($q_i(t)= R_i(t)$) and output ($q_{i+1}(t)= S_i(t)$) vehicle flow of the road cell is represented by corresponding controllable current sources  $I_i(t)$ and $I_{i+1}(t)$, respectively. Now, from Fig.~\ref{fig:concentration_diagram}(b), we observe that current passing through the capacitor $C_i$ is given by
\begin{align}
    C_i \frac{dV_i(t)}{dt} = I_i(t) - I_{i+1}(t) \label{eq:concentration_constraint1}
\end{align}
Using $C_i=1F$ and the similarities between electric variables and dynamic traffic variables, we can reformulate \eqref{eq:concentration_constraint1} as follows:
\begin{align}
    \frac{dn_i(t)}{dt} = q_i(t) - q_{i+1}(t) \label{eq:concentration_constraint2}
\end{align}

\begin{figure}[t]
    \centering
    \includegraphics[width=1\linewidth]{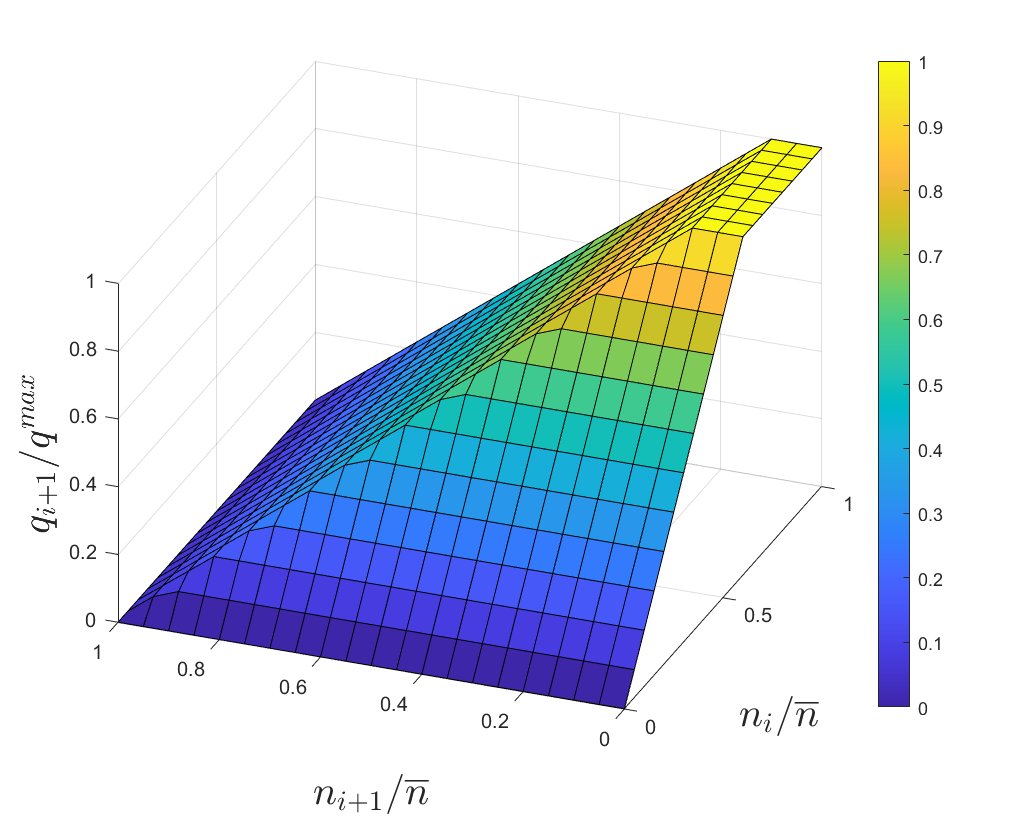}
    \caption{Value of the  controllable current source $q_{i+1}(n_{i}, n_{i+1})$ as function of the normalized cell concentrations $n_{i}$ and $n_{i+1}$. Note: the function inputs and outputs are normalized with respect to their maximum physical limits ($\overline{n}$, $q^{max}$). Parameters: $u^{free}=1$ km/min, $\Delta x=1$ km, $q^{max}$= 6 cars/min, $w=(2/3)\;u^{free} $, $\overline{n}=10$ cars.}
    \label{fig:currentSource}
\end{figure}

Fig.~\ref{fig:flow_diagram} displays the vehicle flow between two cells and the corresponding ECM representation, where the vehicle flow is replaced with a variable current source ($I_{i+1}$). The current value from the controllable current source depends upon the voltage values $V_i(t)$ and $V_{i+1}(t)$. We define this relation as 
\begin{align}
    I_{i+1}(t) := \min \left\{ \gamma_1V_i(t),I^{max}, \gamma_2(\overline{V}-V_{i+1}(t))\right\}\label{eq:flow_constraint1}
\end{align}
where $I^{max}, \gamma_1, \gamma_2 >0$ are some constants and $\overline{V}$ denotes the maximum voltage difference that can be sustained across the capacitor before electrical breakdown occurs. Again, using the similarities between electric variables and dynamic traffic variables, we can reformulate \eqref{eq:flow_constraint1} as follows:
\begin{align}
    q_{i+1}(t) = \min \left\{ \frac{u^{free}}{\Delta x}n_i(t),q^{max}, \frac{w}{\Delta x} (\overline{n}-n_{i+1}(t))\right\}\label{eq:flow_constraint2}
\end{align}
where $\gamma_1 = u^{free}/ \Delta x,\; \gamma_2=w/ \Delta x,\; I^{max}=q^{max}$ and $\overline{n}$ represents the maximum number of vehicles that any cell in the road section can handle with congestion. Fig.~\ref{fig:currentSource} provides the plot of $q_{i+1}(t)$ as a function $n_i(t)$ and $n_{i+1}(t)$. 
\begin{Proposition}
    Given Assumption~\ref{assume:const_density}, the ECM described by \eqref{eq:concentration_constraint2} and \eqref{eq:flow_constraint2} is equivalent to the spatially discretized form of the LWR traffic model given by \eqref{eq:dta_fund_eq}--\eqref{eq:q_rho_relation}, with the fundamental diagram represented by the trapezoidal relationship defined in \eqref{eq:fund_diag}. 
\end{Proposition}
\textbf{Proof:}
Let us first consider \eqref{eq:dta_conti_eq}. Integrating it over cell $i$ (see Fig.~\ref{fig:concentration_diagram}) gives us
\begin{align}
    \int_{x_{i-0.5}}^{x_{i+0.5}} \frac{d \rho}{dt}dx + \int_{x_{i-0.5}}^{x_{i+0.5}} \frac{dq}{dx}dx = 0 \label{eq:cont_part0}.
\end{align}
Now, utilizing \eqref{eq:approx}, the first term becomes 
\begin{align}
    \frac{d}{dt}\int_{x_{i-0.5}}^{x_{i+0.5}} \rho(x,t)dx = \frac{dn_i(t)}{dt} \label{eq:cont_part1}.
\end{align}
The second term follows from the fundamental theorem of calculus:
\begin{align}
    \int_{x_{i-0.5}}^{x_{i+0.5}} \frac{dq}{dx} dx = q(x_{i+0.5},t) - q(x_{i-0.5},t) \label{eq:cont_part2}.
\end{align}
Therefore, from \eqref{eq:cont_part0}, \eqref{eq:cont_part1} and \eqref{eq:cont_part2}, we get
\begin{align}
    \frac{dn_i(t)}{dt} = q(x_{i-0.5},t) - q(x_{i+0.5},t) \label{eq:cont_part3}
\end{align}
Now, it can be observed that $q(x_{i-0,5},t)$ and $q(x_{i+0,5},t)$ are the inflow and outflow values of cell $i$, respectively. Therefore, we can rewrite \eqref{eq:cont_part3} as
\begin{align}
    \frac{dn_i(t)}{dt} = q_{i}(t) - q_{i+1}(t). \label{eq:cont_part4}
\end{align}
So the capacitor balance equation is simply the spatially discretized LWR conservation law. Next, from Fig.~\ref{fig:flow_diagram}, we observe that the upstream cell cannot send vehicles faster than either its free-flow demand or the road capacity. Hence, the amount of traffic that cell $i$ can send is determined by the free-flow side of the trapezoidal fundamental diagram \cite{daganzo1994cell}. Therefore,
\begin{align}
    S_i = \min\{u^{free}\rho_i,q^{max}\} = \min\left\{\frac{u^{free}}{\Delta x}n_i,q^{max}\right\}
\end{align}
where $n_i = \rho_i \Delta x$. Similarly, the downstream cell also imposes a limitation and cannot receive more than its capacity. Hence, the amount of traffic that cell $i+1$ can receive is determined by the congested branch of the trapezoidal fundamental diagram. Therefore,
\begin{align}
    R_{i+1} = \min\{q^{max},w(\rho^{jam}-\rho_{i+1})\} \nonumber\\
    = \min\left\{q^{max},\frac{w}{\Delta x}(\overline{n}-n_{i+1})\right\}
\end{align}
where $\overline{n}=\rho^{jam}\Delta x$ and $n_{i+1}=\rho_{i+1}\Delta x$. Now the actual flow is the minimum of sending and receiving capabilities. Therefore,
\begin{align}
    q_{i+1}(t) &= \min\{S_i,R_{i+1}\} \nonumber \\
    &= \min\left\{\frac{u^{free}}{\Delta x}n_i(t),q^{max},\frac{w}{\Delta x}(\overline{n}-n_{i+1}(t))\right\}
\end{align}
This concludes the proof.

\hspace*{8.5cm}\rule{0.25cm}{0.25cm}
\subsection{Network Input Flow}
Let us consider the input vehicle flow to the network as $q_{in}$, which enters the road link through its upstream cell ($i=0$). Two input-flow scenarios are considered: i) free-flow conditions and ii) restricted-flow conditions. When $n_0=0$, i.e., the upstream cell is empty, and $q_{in}\leq q^{max}$, the entire input flow can enter the road link, yielding $R_0=q_{in}$. This case is referred to as the free-flow condition because the incoming vehicles can enter the road link without restriction.

\begin{figure}[ht]
    \centering
    \includegraphics[width=0.97\linewidth]{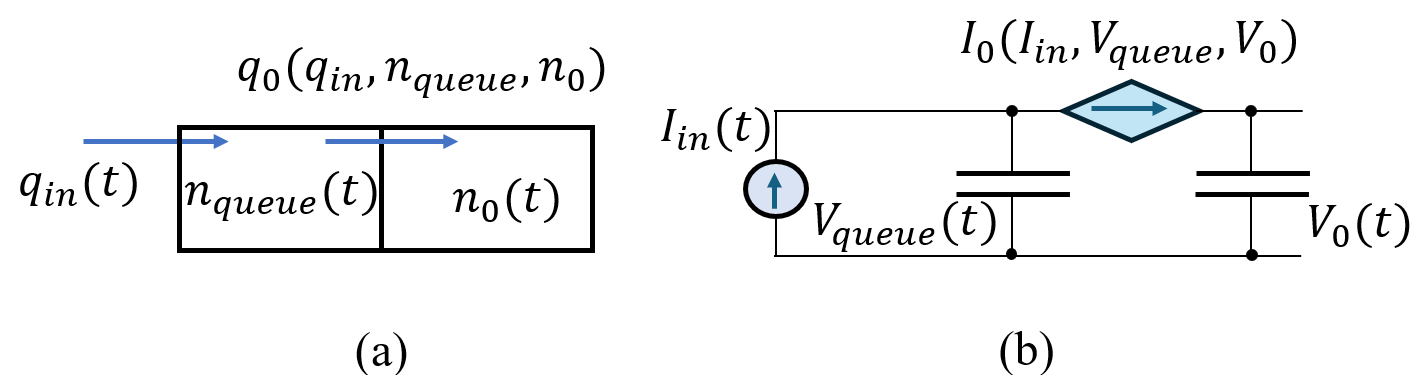}
    \caption{Illustration of (a) inclusion of a queue cell to accommodate excess input flow and (b) the corresponding ECM. }
    \label{fig:queue_cell}
\end{figure}

In contrast, when $q_{in}>q^{max}$, or when $n_0\neq 0$ and $q_{in}>w(\overline{n}-n_0)$, the entire input flow cannot be accommodated by the upstream cell. This case is referred to as the restricted-flow condition. To account for the excess inflow under such conditions, an external queue cell, denoted by $n_{queue}$, is introduced with a sufficiently large vehicle storage capacity ($\overline{n}_{queue}\gg 0$). In the ECM representation, this queue cell is modeled as a capacitor with a correspondingly large breakdown voltage ($\overline{V}{queue}\gg 0$), as illustrated in Fig.~\ref{fig:queue_cell}. 
Therefore, to account for the restricted input flow conditions, we modify the equation for $q_0(t)$ as follows:
\begin{dmath}
    q_0(t) := \min\left\{q_{in}+\max\left(0,n_{queue}u^{free}\right),q^{max},w(\overline{n}-n_0) \right\}  \label{eq:queue}
\end{dmath}
where the term $\max\left(0,n_{queue}u^{free}\right)$ allows us to discharge the queue capacity once the upstream condition becomes uncongested.
\begin{remark}
Similar to the modification of $q_0(t)$, the output flow from the last cell ($i=N$) of a road link can be defined as
\begin{align}
q_N(t) := \min\left\{ n_{N-1}(t)u^{free}, q^{max}, q_{out}(t)\right\}, \label{eq:outflow}
\end{align}
where $q_{out}(t)$ denotes the maximum flow that can be received by the upstream cell of the subsequent road link. If the current road link terminates at an exit from the transportation network, $q_{out}(t)$ is assumed to be unbounded, i.e., $q_{out}(t)\rightarrow\infty$, indicating that the outflow is not constrained by a downstream receiving capacity.
\end{remark}
\begin{figure}[ht]
    \centering

    \begin{subfigure}[b]{0.47\textwidth}
        \centering
        \includegraphics[width=\textwidth]{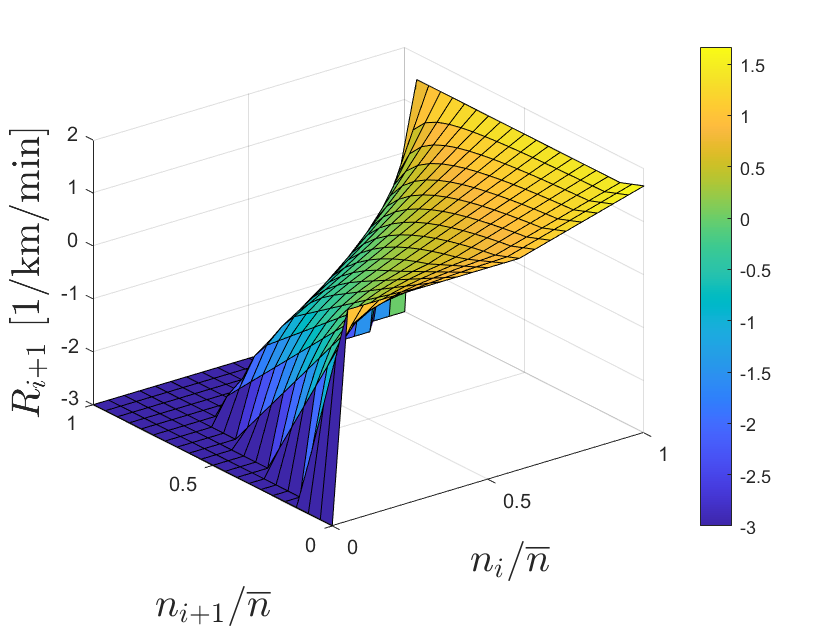}
        \caption{Side view}
        \label{fig:resist_side_view}
    \end{subfigure}
    \hfill
    \begin{subfigure}[b]{0.47\textwidth}
        \centering
        \includegraphics[width=\textwidth]{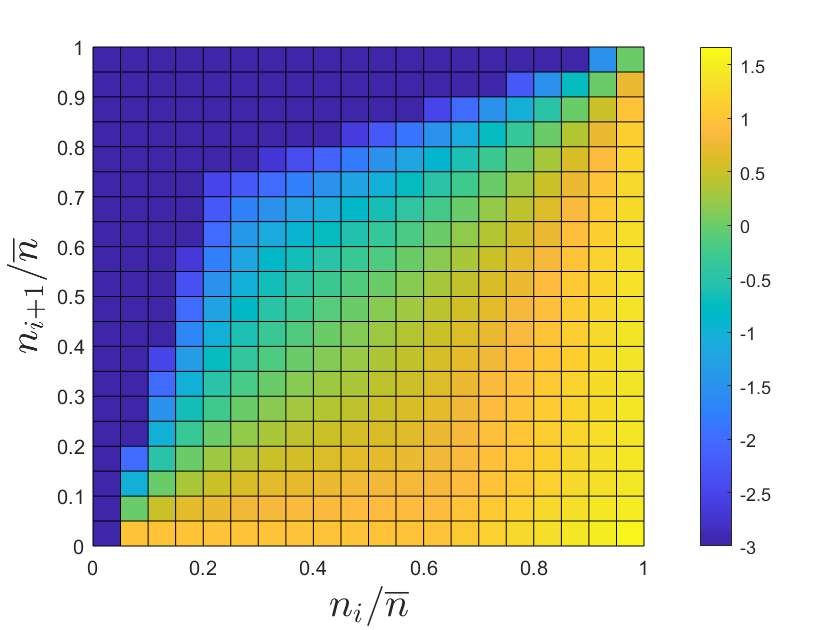}
        \caption{Top view}
        \label{fig:resist_top_view}
    \end{subfigure}

    \caption{Value of the  nonlinear resistance $R_{i+1}(n_{i}, n_{i+1})$ as function of the normalized cell concentrations $n_{i}$ and $n_{i+1}$. Note: the function inputs and outputs are normalized with respect to their maximum physical limits ($\overline{n}$, $q^{max}$). Parameters: $u^{free}=1$ km/min, $\Delta x=1$ km, $q^{max}$= 6 cars/min, $w=(2/3)\;u^{free} $, $\overline{n}=10$ cars. }
    \label{fig:traffic_resistance}
\end{figure}
\subsection{ ECM realization of Resistance to Vehicle Flow}
In this section, we replace the controllable current source described in Section~\ref{section:ECM_proposed} with a nonlinear variable resistance (see Fig.~\ref{fig:resistance_diagram}). Its value is defined as follows:
\begin{figure}[ht]
    \centering
    \includegraphics[width=0.97\linewidth]{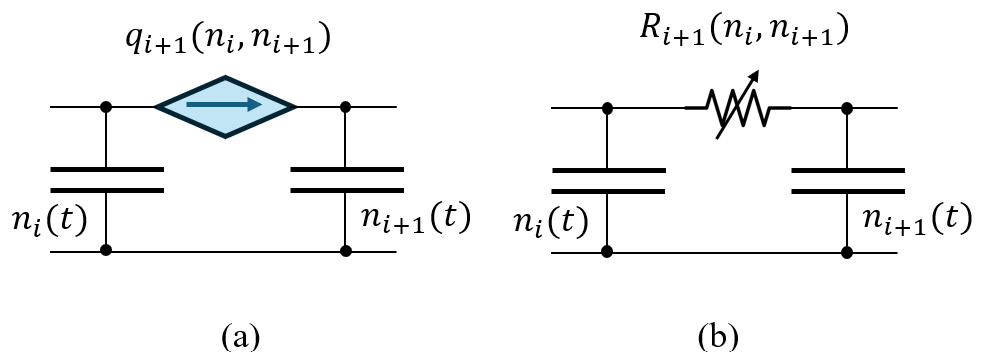}
    \caption{Illustration of (a) ECM with current source and (b) ECM with nonlinear resistance. }
    \label{fig:resistance_diagram}
\end{figure}

\begin{align}
    R_{i+1}(t) := \frac{n_i(t)-n_{i+1}(t) }{ \min \left\{ u^{free}n_i(t),q^{max}, w (\overline{n}-n_{i+1}(t))\right\}} \label{eq:resist_constraint2}
\end{align}
with units of $1/km/min$. Fig.~\ref{fig:traffic_resistance} showcases the different values of $R_{i+1}(t)$ based on \eqref{eq:resist_constraint2}. It can be observed that the resistance can have three types of behaviour:
\begin{itemize}
        \item \textbf{Positive resistance} ($ n_{i} > n_{i+1}$): In this case, the upstream cells have a higher concentration than the downstream cells, and the  ECM behaves as a cascade of RC elements, filtering the input signal. 
        \item \textbf{Zero resistance} ($n_i = n_{i+1}$): In this case, the ECM behaves (potentially temporarily) as a parallel of capacitors.
        \item \textbf{Negative resistance} ($n_i < n_{i+1}$): This corresponds to the case where the downstream has a higher concentration than upstream. This leads to an RC circuit with negative resistance \cite{negative_resistance}.
    \end{itemize}

The ECM-based representation of resistance to vehicle flow provides useful insights into how traffic dynamics evolve toward steady-state conditions. In the case of positive resistance ($n_i>n_{i+1}$), the upstream cell contains more vehicles than the downstream cell, resulting in a higher sending capability. Consequently, vehicles flow from the upstream cell to the downstream cell, tending to reduce the difference in vehicle concentrations between the two cells. Conversely, in the case of negative resistance ($n_i<n_{i+1}$), the downstream cell contains more vehicles than the upstream cell, thereby restricting its receiving capability and reducing the flow between the cells. This restriction promotes vehicle accumulation in the upstream cell, which may subsequently reduce the concentration difference between the two cells. Thus, the ECM representation illustrates how inter-cell flow adjustments can facilitate the redistribution of vehicles and the evolution of traffic toward steady-state conditions.

\subsection{Steady-State Analysis}\label{section:steady-state}
\begin{figure}[t]
    \centering
    \includegraphics[width=0.97\linewidth]{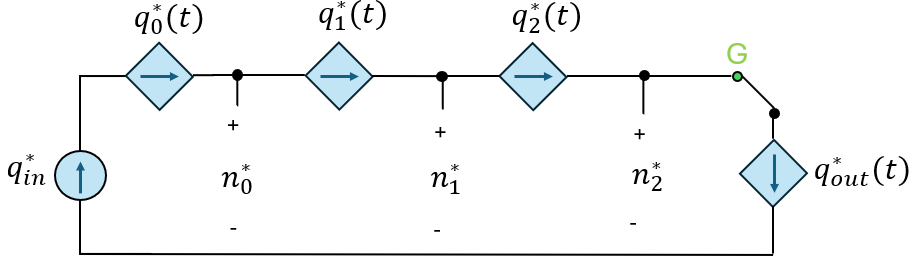}
    \caption{ECM of the road segment with a traffic light under steady-state conditions when the traffic light is green.}
    \label{fig:ttraffic_light_steady_state}
\end{figure}
Fig.~\ref{fig:ttraffic_light_steady_state} presents the ECM of the road segment with a traffic light under steady-state conditions when the traffic light is green (to simplify the presentation, we ignore the queue capacitor). In steady-state conditions, the capacitors have constant voltage and can be approximated as "open circuits".
In this case, the steady-state circuit is composed of a series of current sources. To obtain a valid equilibrium point, all the (input and controllable) current sources need to have the same value due to their series connection. This leads to the following  set of stationary conditions for the ECM, obtained by setting $d n_i/dt =0$ in ~\eqref{eq:concentration_constraint1}:
\begin{align}
   0 &= q_0^*(t) - q_1^*(t) \nonumber \\
   0 &= q_1^*(t) - q_2^*(t) \nonumber \\
   0 &= q_2^*(t) - q_{out}^*(t) \nonumber
\end{align}
where the notation $q_i^*$ denotes the equilibrium point for $q_i$.

\textbf{Regime I (No congestion)}:
In this regime, we assume that the steady-state current sources $q^*_{in}$ are not congested.
Note that all the controlled current sources introduced through \eqref{eq:flow_constraint2} are determined based on the minimum of three quantities ($x,q^{max},z$):
\begin{eqnarray}
    q_{in} (.)=  \min\{x, q^{max}, z\}\label{eq:currentSource_general}
\end{eqnarray}
where the term $x$ is related to a variable "upstream" ($i$)  in the circuit and  the term $z$ is related to what happens at the cell $i+1$ or downstream the circuit (e.g. $q_{out}$). We say that the current sources are non-congested if $x< \min\{q^{max},z\}$ in ~\eqref{eq:currentSource_general}. Assuming that all current sources operate in the non-congested state, we have
\begin{eqnarray}
    q^*_{in} = n^*_{i} u^{free} = n^*_{out} u^{free}, \qquad q^*_{in} < R^*_{out}
\end{eqnarray}
for $i=0,1,2$. Here, $R^*_{out}$ denotes the input vehicle flow that can be accommodated by the upstream cell of the next road section. In this case, the input source ($q^*_{in}$) sets the current in the controllable current sources and in the circuit as long as the output source can cope with the input flow   (i.e., $q^*_{in} < R^*_{out}$).

\textbf{Regime II (Congested):}
In this regime, we assume that all steady-state current sources ($q_i^*$) are congested, i.e., $z<\min\{x,q^{max}\}$ in \eqref{eq:currentSource_general}, leading to
\begin{eqnarray}
    R^*_{out}=q^*_{out} = 
      w (\overline{n}_i - n^*_i )  = q^*_{in}            
\end{eqnarray}
for $i=0,1,2$.
In this regime, it is the output source $R^*_{out}$ that sets the current in the overall circuit.
\section{Validation through SUMO}\label{section:results}
\begin{figure}[t]
    \centering
    \includegraphics[width=0.97\linewidth]{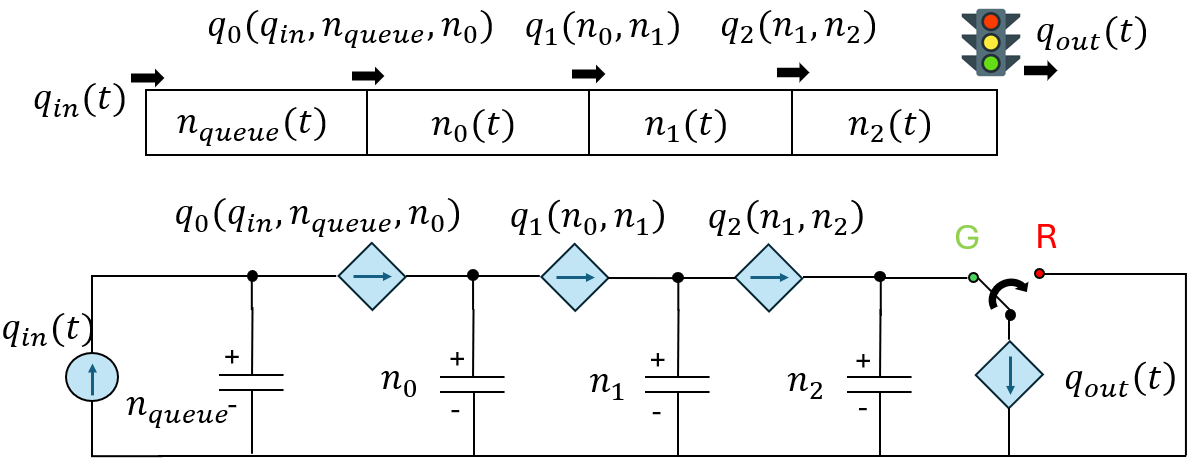}
    \caption{Equivalent circuit model of a road segment discretized with $N=3$ cells, using controllable current sources between capacitors.}
    \label{fig:traffic_stop_example}
\end{figure}
In this work, Simulation of Urban MObility (SUMO) is employed as a traffic simulation environment to validate the proposed ECM. SUMO is particularly suitable for validating dynamic traffic models under a variety of operating conditions. It is an open-source, microscopic traffic simulation platform widely used for modeling and analyzing transportation networks \cite{Behrisch2011}. SUMO represents the motion and interactions of individual vehicles while allowing users to define road-network geometry, traffic demand, vehicle characteristics, routes, traffic-control devices, and other mobility-related parameters. Its microscopic representation enables traffic quantities such as vehicle flow, density, speed, queue formation, and congestion propagation to be evaluated at different locations and time instances.

\begin{table}[t]
\centering
\caption{Simulation parameters for ECM and SUMO validation.}
\label{tab:simulation_parameters}
\resizebox{\columnwidth}{!}{%
\begin{tabular}{lcccc}
\toprule
\textbf{Parameter} & \textbf{Symbol}
& \textbf{Traffic Light} & \textbf{Merge}
& \textbf{Diverge} \\
\midrule
Number of road segments
& -- & 1 & 3 & 3 \\

Cells per road segment
& $N$ & 3 & 3 & 3 \\

Cell length (km)
& $\Delta x$ & 1 & 1 & 1 \\

Maximum concentration (veh/km)
& $\bar{n}$ & 120 & 120 & 120 \\

Free-flow speed (km/h)
& $u^{free}$ & 45 & 45 & 45 \\

Backward wave speed (km/h)
& $w$ & 16.2 & 16.2 & 16.2 \\

Input flow (veh/min)
& $q_{in}$ & 6, 18, 30 & 40 (each) & 12 \\

Simulation duration (min)
& $T$ & 22 & 60 & 60 \\

ECM capacitance (F)
& $C_i$ & 1 & 1 & 1 \\

\bottomrule
\end{tabular}%
}
\end{table}

\textbf{Traffic Analysis at Traffic Light:}

\begin{figure}[htbp]
    \centering

    \begin{subfigure}[b]{0.47\textwidth}
        \centering
        \includegraphics[width=\textwidth]{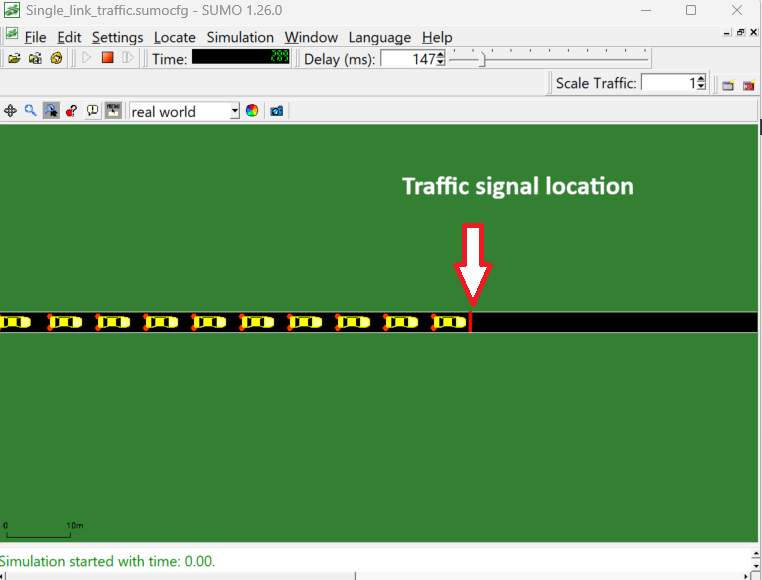}
        \caption{Traffic signal is RED.}
        \label{fig:SUMO_traffic_stop}
    \end{subfigure}
    \hfill
    \begin{subfigure}[b]{0.47\textwidth}
        \centering
        \includegraphics[width=\textwidth]{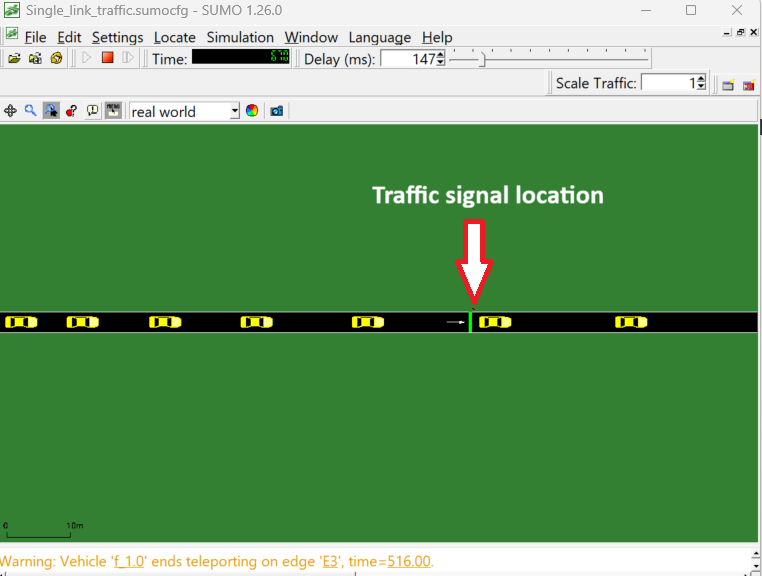}
        \caption{Traffic signal is GREEN.}
        \label{fig:SUMO_traffic_go}
    \end{subfigure}

    \caption{SUMO simulations involving a traffic stop.}
    \label{fig:SUMO_traffic}
\end{figure}

Let us consider a single road segment with a traffic signal located at its downstream end, as illustrated in Fig.~\ref{fig:traffic_stop_example}. The road segment is discretized into three cells ($N=3$), each with a length of $1$ km. Also, $\overline{n}=120$ vehicle/km, $u^{free}=45$ km/h, and $w=0.36\; u^{free}$ km/h. In the corresponding ECM, the traffic signal is represented by a switch that controls the downstream current source $q_{out}(t)$. During a red-light phase, the switch is opened, thereby preventing current from leaving the circuit and, equivalently, setting the downstream traffic outflow to zero. During a green-light phase, the switch is closed, allowing current to flow through $q_{out}(t)$ and enabling vehicles to exit the downstream cell subject to the applicable flow constraints. This switching mechanism therefore captures the alternating restriction and release of traffic flow imposed by the downstream traffic signal. In the test example considered in this paper, the entire simulation is run for 22 minutes, out of which the red signal is activated for 11 minutes and the remaining time is allocated for the yellow and green signals. In this experiment, the yellow signal is considered the same as the green signal. Fig.~\ref{fig:SUMO_traffic} illustrates the SUMO traffic-signal simulation during the red and green phases. To quantify the discrepancy between the proposed ECM and SUMO simulations, the percentage error is calculated as the absolute difference between their predicted vehicle concentrations, normalized by the maximum reference concentration and multiplied by 100.
\begin{equation}
e_i(t) =
\frac{100\left|n_i^{\mathrm{SUMO}}(t)
-n_i^{\mathrm{ECM}}(t)\right|}{n_{\max}},\label{eq:error_calc}
\end{equation}
where $n_i^{\mathrm{SUMO}}(t)$ and
$n_i^{\mathrm{ECM}}(t)$ represent the vehicle concentrations
in cell $i$ at time $t$, obtained from SUMO and the
proposed ECM, respectively. Here, $n_{\max} = \max_{i,t}n_i^{\mathrm{SUMO}}(t)$ denotes
the maximum reference vehicle concentration used for
normalization.

 During the red phase, vehicles accumulate upstream of the signal and form a queue, while during the green phase, the queued vehicles are released and traffic begins to discharge.
\begin{figure}[htbp]
    \centering

    \begin{subfigure}[b]{0.49\textwidth}
        \centering
        \includegraphics[width=\textwidth]{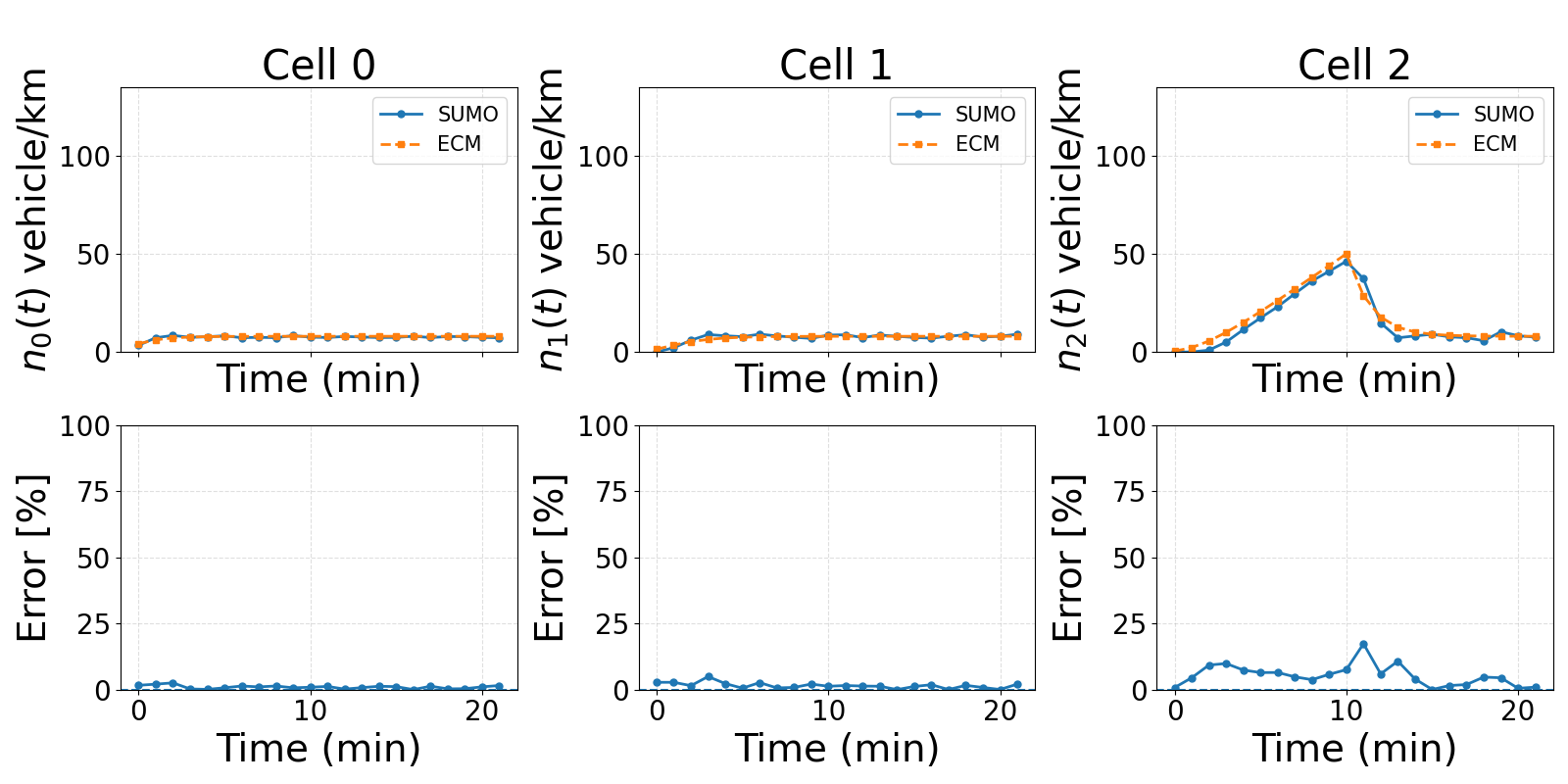}
        \caption{$q_{in}=6$ vehicles/min}
        \label{fig:input6}
    \end{subfigure}
    \hfill
    \begin{subfigure}[b]{0.49\textwidth}
        \centering
        \includegraphics[width=\textwidth]{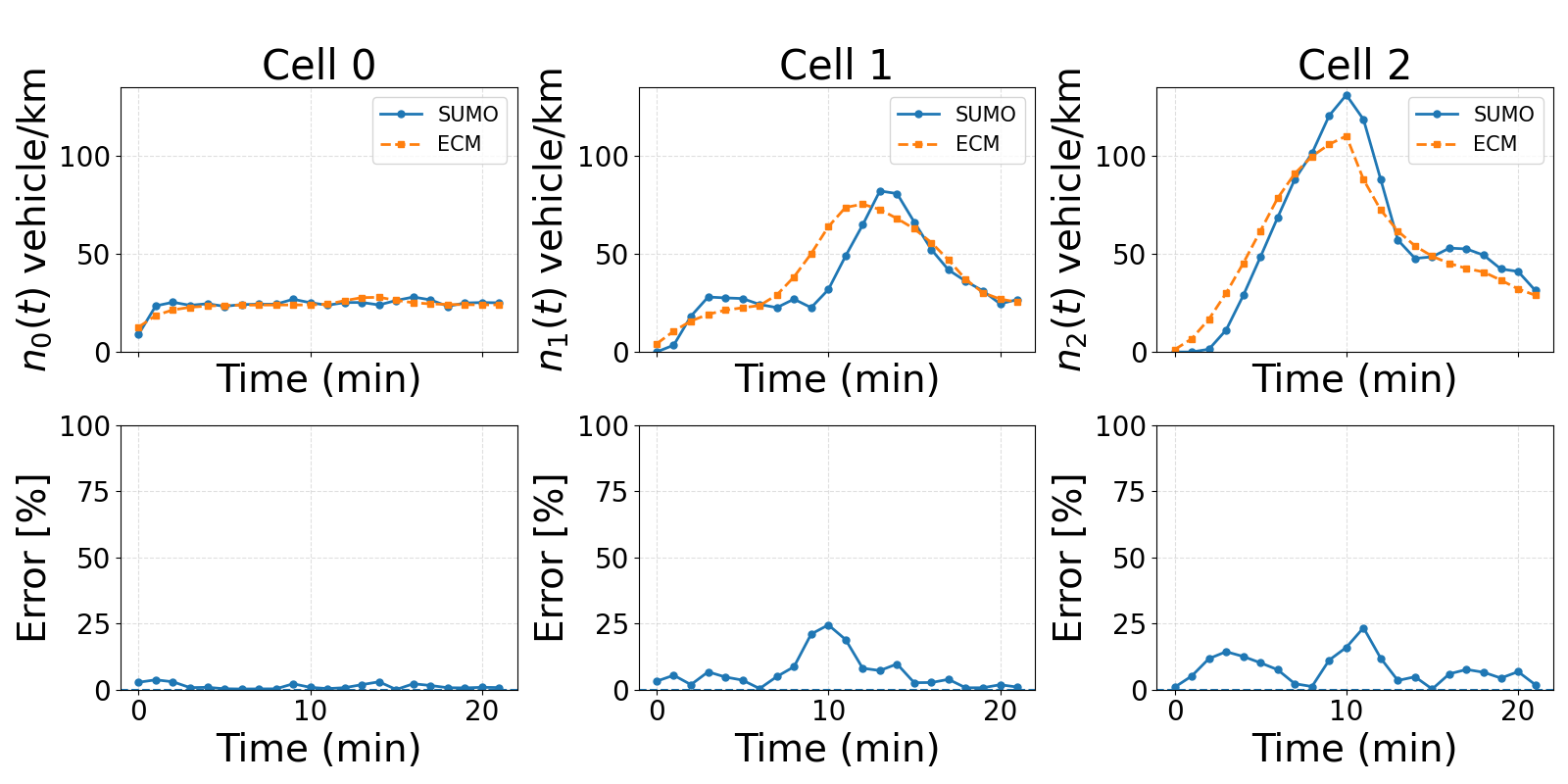}
        \caption{$q_{in}=18$ vehicles/min}
        \label{fig:input18}
    \end{subfigure}
    \hfill
    \begin{subfigure}[b]{0.49\textwidth}
        \centering
        \includegraphics[width=\textwidth]{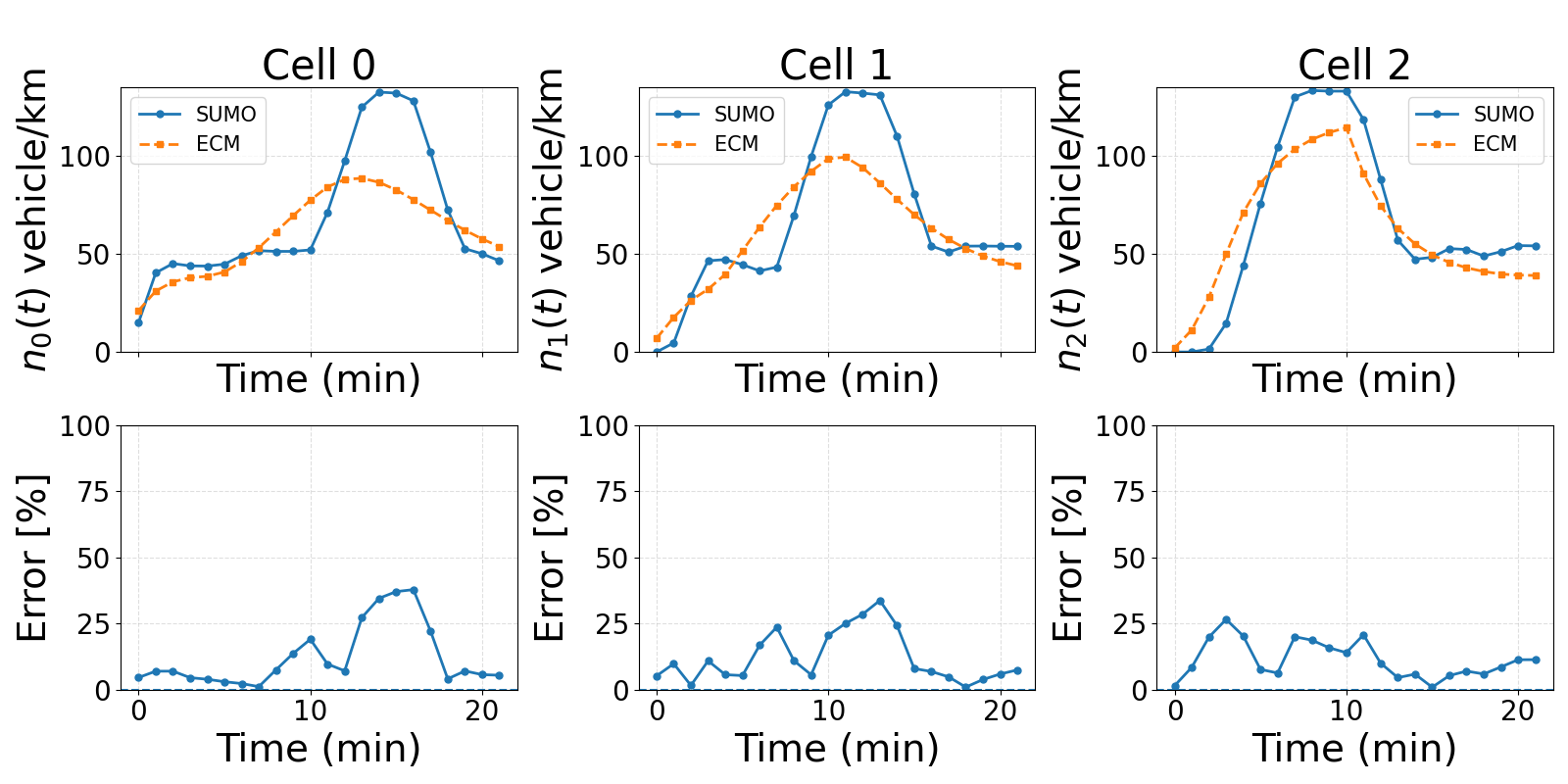}
        \caption{$q_{in}=30$ vehicles/min}
        \label{fig:input30}
    \end{subfigure}

    \caption{Comparison of vehicle concentration values for different road cells obtained from the SUMO simulation and ECM optimization, along with the corresponding concentration errors in a traffic light simulation.}
    \label{fig:traffic}
\end{figure}

Figs.~\ref {fig:input6}-\ref{fig:input30} provide the validation results of the proposed ECM against the microscopic SUMO simulation under progressively heavier input vehicle flow conditions. All three figures use the same three-cell road segment with a downstream traffic signal. Fig.~\ref{fig:input6} represents the lowest-demand case with $q_{in}=6$ vehicle/min. The concentration histories show that the ECM follows the general temporal behavior predicted by SUMO quite closely, particularly in Cells 0 and 1. The concentrations in these upstream cells remain comparatively low and vary only moderately during the simulation. The most significant change occurs in Cell 2, the downstream cell adjacent to the traffic signal. During the red-light period, vehicles cannot leave the road segment, so vehicles accumulate in Cell 2. The concentration therefore rises substantially before decreasing after the signal allows traffic to leave. The ECM reproduces the buildup and subsequent discharge of this downstream queue quite well. The largest discrepancy between SUMO and ECM in Fig.~\ref{fig:input6} appears near the transient peak and during the rapid reduction in Cell 2 concentration. The corresponding error plot therefore exhibits its largest error around this transition. In contrast, the errors in Cells 0 and 1 remain relatively small.

Fig.~\ref{fig:input18} shows a substantially different traffic regime. With $q_{in}$ increased from 6 to 18 vehicles/min, congestion is no longer confined primarily to Cell 2. The queue created by the downstream traffic signal begins to propagate upstream into Cell 1, and to a lesser extent toward Cell 0. Cell 2 again experiences the earliest and strongest concentration increase because it is immediately upstream of the traffic signal. The concentration rises rapidly while the signal restricts the downstream outflow. Cell 1 subsequently develops a pronounced buildup, reflecting the backward propagation of congestion from the downstream end of the road. The errors in Fig.~\ref{fig:input18} are larger than those in Fig.~\ref{fig:input6}, especially in Cell 1. The most pronounced deviations occur during the growth and dissipation of congestion. The ECM vehicle concentration tends to be smoother and exhibits somewhat different transition timing compared with SUMO. That difference is expected because the ECM treats each cell as having a uniform concentration, whereas SUMO explicitly models heterogeneous vehicle positions within each cell. Another notable feature is that after the concentration reaches its maximum, SUMO and ECM dissipate the accumulated traffic at a similar rate.

Fig.~\ref{fig:input30} represents the most congested case and provides the most challenging test of the ECM. At $q_{in}=30$ vehicles/min, the traffic demand is much greater than in Fig.~\ref{fig:input6} and \ref{fig:input18}. As a result, congestion initiated at the downstream traffic signal propagates through all three cells. Cell 2 fills rapidly because of the downstream restriction. As Cell 2 becomes congested, Cell 1 therefore begins to accumulate vehicles. Eventually, the congestion propagates further upstream, and Cell 0 also exhibits a large increase in concentration. Fig.~\ref{fig:input30} also reveals the largest differences between SUMO and ECM. The error curves for Cells 0 and 1 are considerably more pronounced than in the lower-inflow cases. The ECM captures the broad buildup, saturation, and recovery pattern, but the timing and magnitude of some concentration transitions differ from SUMO. One particularly visible difference is during the post-peak dissipation phase. SUMO often shows a quicker reduction in concentration, whereas the ECM produces slower discharge. This suggests that the principal source of disagreement is not the fundamental congestion mechanism, but rather the detailed transient dynamics.

The ECM results in Fig.~\ref{fig:traffic} also support the steady-state analysis developed in Section~\ref{section:steady-state}. In each case, the vehicle concentrations across the different cells gradually converge to a common value as time progresses, indicating that the road approaches a spatially uniform steady-state condition. Specifically, the ECM concentrations in Fig.~\ref{fig:input6} converge to approximately 8 vehicles/km, while those in Fig.~\ref{fig:input18} and Fig.~\ref{fig:input30} converge to approximately 24 vehicles/km and 39 vehicles/km, respectively. This convergence is consistent with the theoretical steady-state behavior predicted by the ECM formulation and demonstrates that, after the transient dynamics subside, the vehicle concentrations in all cells approach the corresponding steady-state values.

\section{Traffic Analysis for 2-Lane Merge}\label{section:merge}
\begin{figure}[htbp]
    \centering

    \begin{subfigure}[b]{0.49\textwidth}
        \centering
        \includegraphics[width=\textwidth]{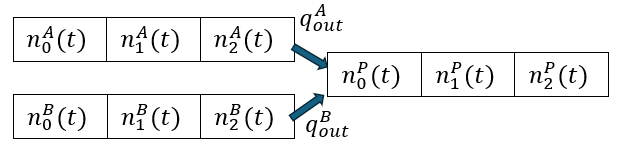}
        \caption{Two-lane merge example.}
        \label{fig:road_merge}
    \end{subfigure}
    \hfill
    \begin{subfigure}[b]{0.4\textwidth}
        \centering
        \includegraphics[width=\textwidth]{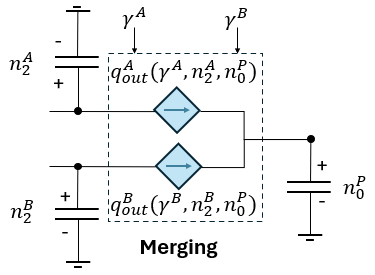}
        \caption{ECM representation of road $A$ and $B$ merging into road $P$.}
        \label{fig:merge}
    \end{subfigure}
    
    \caption{Illustration of a road-merge example, where two single-lane roads, $A$ and $B$, merge into a single-lane downstream road, $P$.}
    \label{fig:merge_presentation}
\end{figure}

In this section, we model the case of two single-lane roads merging into a single-lane downstream road, as shown in Fig.~\ref{fig:road_merge}. Each road segment is discretized into three cells ($N=3$), each of length $1$ km. In this scenario, the three roads share identical traffic parameters, namely, $\overline{n}=120$ vehicle/km, $u^{free}=45$ km/h, and $w=0.36\,u^{free}$ km/h. Also, $q_{in}^A(t)=q_{in}^B(t)= 40$ vehicles/min. We can construct the ECM for each road in the same way as in the traffic-light example discussed in Section~\ref{section:results}. Next, we design an ECM to represent the merging of vehicle flows from roads $A$ and $B$ into road $P$, as illustrated in Fig.~\ref{fig:merge}. Specifically, we connect the downstream cells of roads $A$ and $B$, represented by their corresponding capacitors, to the upstream cell of road $P$ through controllable current sources. The outputs of these current sources represent the vehicle flows $q^A_{out}$ and $q^B_{out}$ exiting roads $A$ and $B$, respectively.

Now, we know that road $P$ needs to be able to accommodate the summation of flows from $A$ and $B$. Therefore,
\begin{align}
    q^A_{out}(t) + q^B_{out}(t) \le \min\left\{q^{max},w(\overline{n}-n^{P}_0\right\}  \label{eq:merge_constraint}
\end{align}
where the right-hand side of \eqref{eq:merge_constraint} represents the maximum vehicle intake allowed by the upstream cell of road $P$ due to its already existing vehicle concentration. In the unconstrained case, we get
\begin{align*}
    q^A_{out}(t) = n^A_2(t)u^{free},\; q^B_{out}(t) = n^B_2(t)u^{free}.
\end{align*}
However, when the combined merge flows are excessive, it becomes necessary to reduce the outgoing flows from the incoming roads. Such a reduction is not unique, as multiple flow-allocation strategies can satisfy the merge constraint in \eqref{eq:merge_constraint}. To mathematically characterize this reduction, we introduce a factor $\beta^{j}\in[0,1]$ for each incoming road $j\in\{A,B\}$, representing the normalized reduction in the flow:
\begin{equation}
   \sum_{j\in\{A,B\}}\beta^{j} n^{j}_2(t) u^{free} 
    \leq \min\left\{\, q_{\max},\, w\left(\overline{n} - n^{P}_0\right) \right\}
\end{equation}
The inclusion of the tunable variables $\beta^{j}$ provides additional degrees of freedom for satisfying the downstream merge constraints. Ideally, we seek to maintain $\beta^{j}$ as close to unity as possible so as to minimize the reduction in the outgoing flow from each incoming road. Mathematically, this objective can be formulated as minimizing:
\[
    J_{1}\!\left(\beta^{j}\right) = \sum_{j\in \{A,B\}}\left(\beta^{j} - 1\right)^{2}.
\]
There may also be scenarios in which one of the incoming roads is assigned a higher priority than the other, for example, due to the presence of a ramp meter, stop sign, or other traffic-control mechanism. In such cases, the relative allocation of the merge flows can be characterized by predefined desired merge ratios $\gamma^{j}$, i.e.,
\begin{align}
    \gamma^{j} = \frac{\beta^{j}}{\sum_{j \in \{A,B\}} \beta^{j}}, \label{eq:gamma_priority}
\end{align}

with $\gamma^{j} \in [0,1]$ and  $\sum \gamma^{j} = 1$. These ratios, which are defined by the user based on the structure of the merge conditions, can be enforced via minimization of the following secondary cost term:  
\[
    J_{2}\!\left(\beta^{j}\right) = \sum_{j \in \{A,B\}}\left(\beta^{j} - \gamma^{j}\left(\sum_{j \in \{A,B\}}\beta^{j}\right)\right)^{2}.
\]
This cost penalizes the violations of the equality given in \eqref{eq:gamma_priority}. The final calculation of the reduction ratio $\beta^{j}$ can be determined via optimization:
\begin{equation}\label{eq:merge_opt}
\begin{aligned}
    & P^{\mathrm{merge}}:
    \quad \min_{\beta^{j}}\; w_{1} J_{1}\!\left(\beta^{j}\right) + w_{2} J_{2}\!\left(\beta^{j}\right) \\
    &\text{s.t.} \quad
     \sum_{j}\beta^{j} n^{j}_2(t) u^{free} 
    \leq \min\left\{\, q_{\max},\, w\left(\overline{n} - n^{P}_0\right) \right\} \\
    & \qquad 0 \leq \beta^{j} \leq 1, \quad j \in \{A, B\}
\end{aligned}
\end{equation}
 where $w_1,w_2$ are weights that assign different priorities to the two cost terms. Ideally, $w_1 \gg w_2 $, in order to maximize the flow of vehicles in the unconstrained case. 
 
\begin{figure}[ht]
    \centering
    \includegraphics[width=0.97\linewidth]{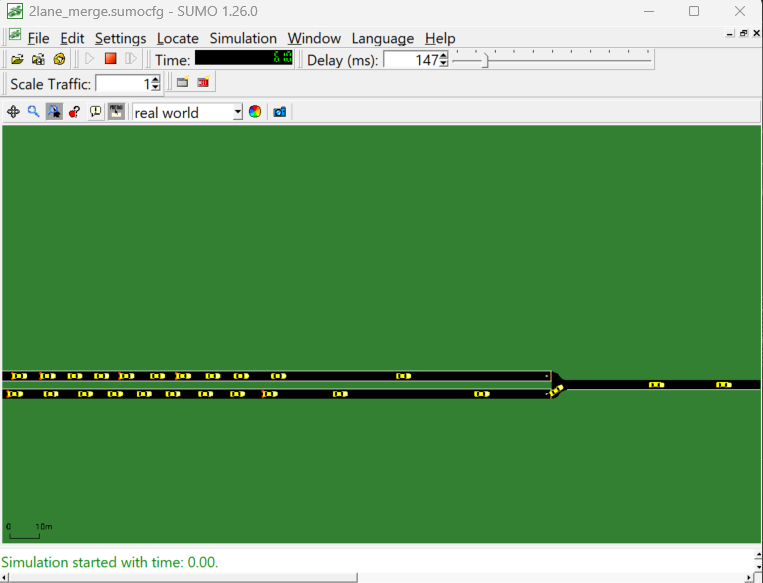}
    \caption{SUMO simulations involving a merge action.}
    \label{fig:SUMO_merge}
\end{figure}

 {\bf Validation in SUMO:}
 Fig.~\ref{fig:SUMO_merge} illustrates the SUMO simulation of the two-road merging scenario, where vehicles traveling on roads $A$ and $B$ converge into the single downstream road $P$. Here, $\gamma^A=\gamma^B=0.5$. The snapshot shows vehicle interactions and accumulation near the merge point as the two incoming traffic streams compete for the available downstream road capacity.

\begin{figure}[htbp]
    \centering

    \begin{subfigure}[b]{0.49\textwidth}
        \centering
        \includegraphics[width=\textwidth]{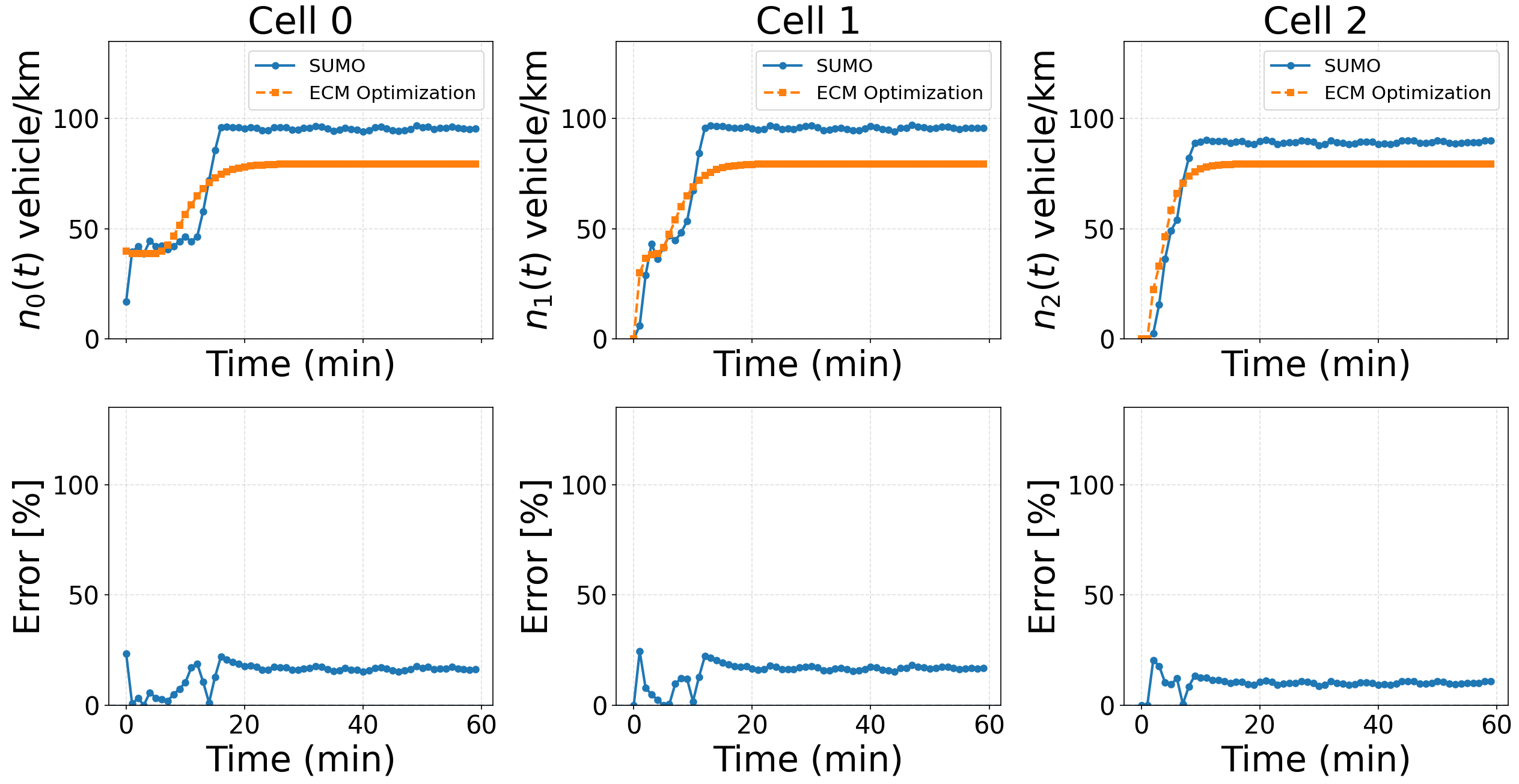}
        \caption{Road $A$}
        \label{fig:mergeA}
    \end{subfigure}
    \hfill
    \begin{subfigure}[b]{0.49\textwidth}
        \centering
        \includegraphics[width=\textwidth]{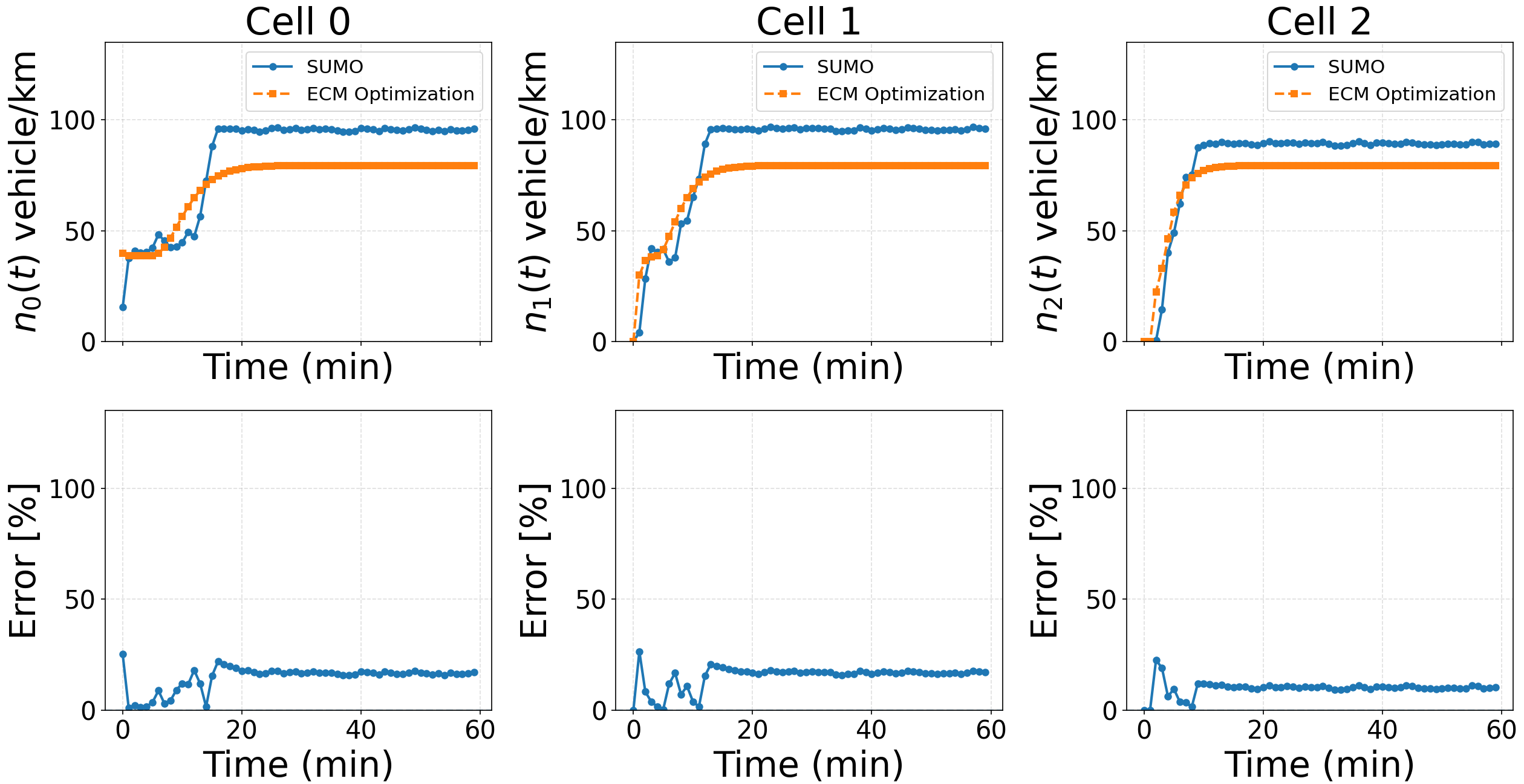}
        \caption{Road $B$}
        \label{fig:mergeB}
    \end{subfigure}
    \hfill
    \begin{subfigure}[b]{0.49\textwidth}
        \centering
        \includegraphics[width=\textwidth]{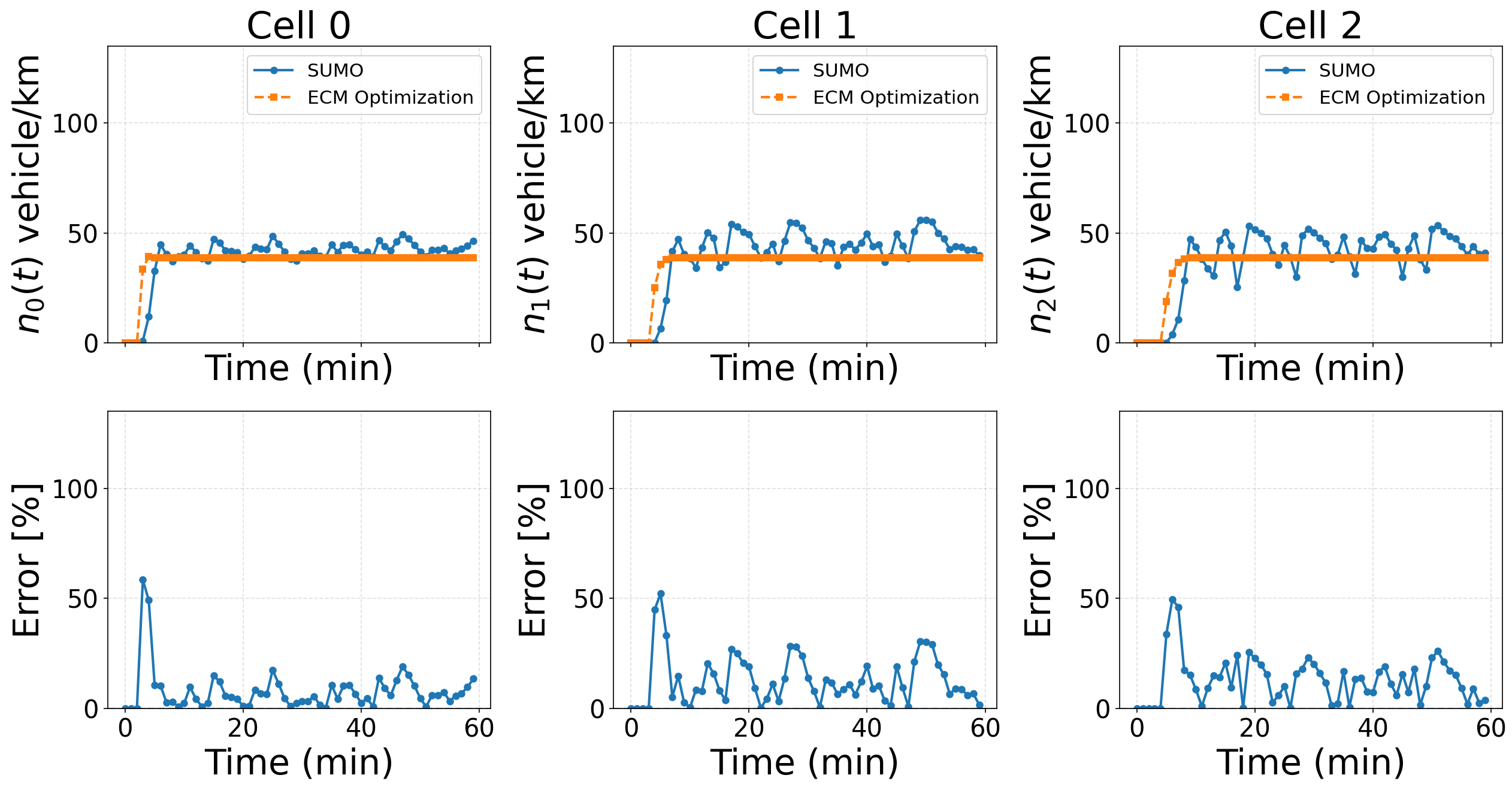}
        \caption{Road $P$}
        \label{fig:mergeP}
    \end{subfigure}

    \caption{Comparison of vehicle concentration values for different road cells obtained from the SUMO simulation and ECM optimization, along with the corresponding concentration errors for the merging simulation.}
    \label{fig:merge_results}
\end{figure}

 Fig.~\ref{fig:mergeA}-\ref{fig:mergeP} provide the validation results for the two-lane merge example by comparing the results of the proposed ECM with those generated by the microscopic SUMO simulation. Fig.~\ref{fig:mergeA}, \ref{fig:mergeB} and \ref{fig:mergeP} corresponds to roads $A$, $B$ and $P$ respectively. Each figure contains three concentration plots on the top row, one for each cell, and three corresponding percentage-error plots on the bottom row. 

Figs. \ref{fig:mergeA} and \ref{fig:mergeB} compare the vehicle concentrations obtained from SUMO and the proposed ECM for Cells 0, 1, and 2 of the incoming roads A and B, respectively. Both roads exhibit similar concentration profiles due to their identical traffic parameters and equal merge priorities $(\gamma^A=\gamma^B=0.5)$. In Cell 0, the concentration initially increases as vehicles enter the road, followed by a rapid rise toward a congested steady-state condition. A similar trend is observed in Cells 1 and 2, where the concentrations gradually increase before stabilizing. This accumulation occurs because the combined traffic demand from roads A and B exceeds the receiving capacity of the downstream road P, restricting the outflow from both incoming roads. Consequently, vehicles accumulate along the incoming roads until the inflow and outflow approach equilibrium. The ECM successfully reproduces the overall transient accumulation and subsequent saturation observed in SUMO across all three cells. However, the steady-state concentrations predicted by the ECM are consistently lower than those obtained from SUMO. This difference can be attributed to the aggregate flow representation used in the ECM, which does not explicitly capture microscopic vehicle interactions near the merging location.

The second row of Figs. \ref{fig:mergeA} and \ref{fig:mergeB} present the percentage errors between the vehicle concentrations predicted by the ECM and SUMO, calculated using \eqref{eq:error_calc}. For both roads, the largest transient errors occur primarily during the initial 10–15 minutes, when vehicle concentrations increase rapidly toward their congested operating conditions. These errors arise mainly from differences in the timing and magnitude of vehicle accumulation predicted by the two models. After the initial transient period, the errors decrease and remain relatively stable, indicating that both models predict similar long-term concentration trends. Nevertheless, a persistent nonzero error remains because the ECM slightly underestimates the steady-state concentrations in all three cells. Overall, the error profiles demonstrate that the proposed ECM captures the dominant congestion dynamics and steady-state behavior of the merging scenario, with discrepancies primarily associated with transient microscopic traffic interactions and differences in equilibrium concentration levels.

Fig.~\ref{fig:mergeP} is qualitatively different from Fig.~\ref{fig:mergeA} and \ref{fig:mergeB}. It gives the concentrations in Cells 0 -- 2 of the downstream receiving road $P$. Road $P$ receives traffic from both upstream roads. Consequently, instead of being fed by a prescribed smooth external inflow, its input is generated dynamically by the interaction of two merging traffic streams. This causes the SUMO concentrations to display substantially more short-time fluctuations. The ECM, by contrast, operates at the cell-aggregated/macroscopic level, so its predictions are considerably smoother. Cell 0 is immediately downstream of the merge. The ECM concentration rises rapidly from its initial condition to an approximately constant value around the first few minutes. SUMO, however, fluctuates noticeably around that level. These fluctuations are physically plausible because individual vehicles from roads $A$ and $B$ do not merge into $P$ as a perfectly continuous flow. Vehicle interactions at the merge produce short-term variations in arrival rate, gaps, acceleration, and lane interaction. The ECM essentially predicts the average macroscopic concentration around which these microscopic SUMO fluctuations occur. The same behavior appears more prominently in Cells 1 and 2. SUMO shows repeated oscillations in concentration throughout the simulation, whereas the ECM produces a much smoother trajectory. Nevertheless, the ECM level generally lies near the center of the SUMO oscillations.
\section{Traffic Analysis for Single Lane Diverge}
\begin{figure}[htbp]
    \centering

    \begin{subfigure}[b]{0.49\textwidth}
        \centering
        \includegraphics[width=\textwidth]{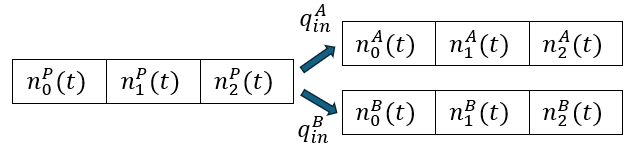}
        \caption{Single-lane diverge example.}
        \label{fig:road_diverge}
    \end{subfigure}
    \hfill
    \begin{subfigure}[b]{0.4\textwidth}
        \centering
        \includegraphics[width=\textwidth]{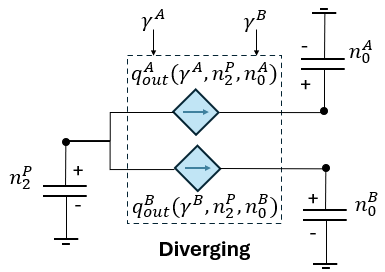}
        \caption{ECM representation of road $P$ diverging into road $A$ and $B$.}
        \label{fig:diverge}
    \end{subfigure}
    
    \caption{Illustration of a road-diverge example, where a single-lane road, $P$, splits into two single-lane downstream roads, $A$ and $B$.}
    \label{fig:diverge_presentation}
\end{figure}
In this section, we develop an ECM to tackle the dual problem of diverging into two roads, $A$ and $B$ (see Fig.~\ref{fig:road_diverge}). We assume that a split ratio $\gamma^{j} \in [0,1]$, $\sum \gamma^{j} = 1$, is defined in advance. For the split ratio to be feasible, we need to fulfill the constraints
\begin{equation}
    \gamma^{j} n_2^P(t)u^{free}
    \leq \min\left\{\, q_{\max},\, w\left(\bar{n} - n^{j}_0 \right) \right\}
\end{equation}
for $j \in J= \{A, B\}$. 
If these constraints are fulfilled, then the downstream roads $A$ and $B$ are able to receive the fraction of traffic coming from the upstream road $P$.
On the other hand, if these constraints are not fulfilled, then we reduce the divergent flows using the factors $\beta^{j} \in [0,1]$:
\begin{equation}
    \beta^{j} \gamma^{j} n_{2}^P u^{free}
    \leq \min\left\{\, q_{\max},\, w\left(\bar{n} - n_{0}^{j} \right) \right\}
\end{equation}
 Similarly to the merge case, $\beta^{j} $ can be computed via optimization:
\begin{equation} \label{eq:diverge_opt}
\begin{aligned}
    P^{\mathrm{diverge}}:&\min_{\beta^{j}}\; J_{1}\!\left(\beta^{j}\right) \\
    &\text{s.t.} \quad
    \beta^{j} \gamma^{j} n_{2}^P u^{free}
      \leq \min\left\{\, q_{\max},\, w\left(\bar{n} - n_{0}^{j} \right) \right\} \\
    & \qquad 0 \leq \beta^{j} \leq 1, \quad j \in \{A, B\}
\end{aligned}
\end{equation}
 
We call $\beta^{j}\!\left(\gamma^{j}, n_{2}^P, n_{0}^{j}\right)$ the solution of this problem, which is then used to parameterize the controllable current source.

\begin{figure}[ht]
    \centering
    \includegraphics[width=0.97\linewidth]{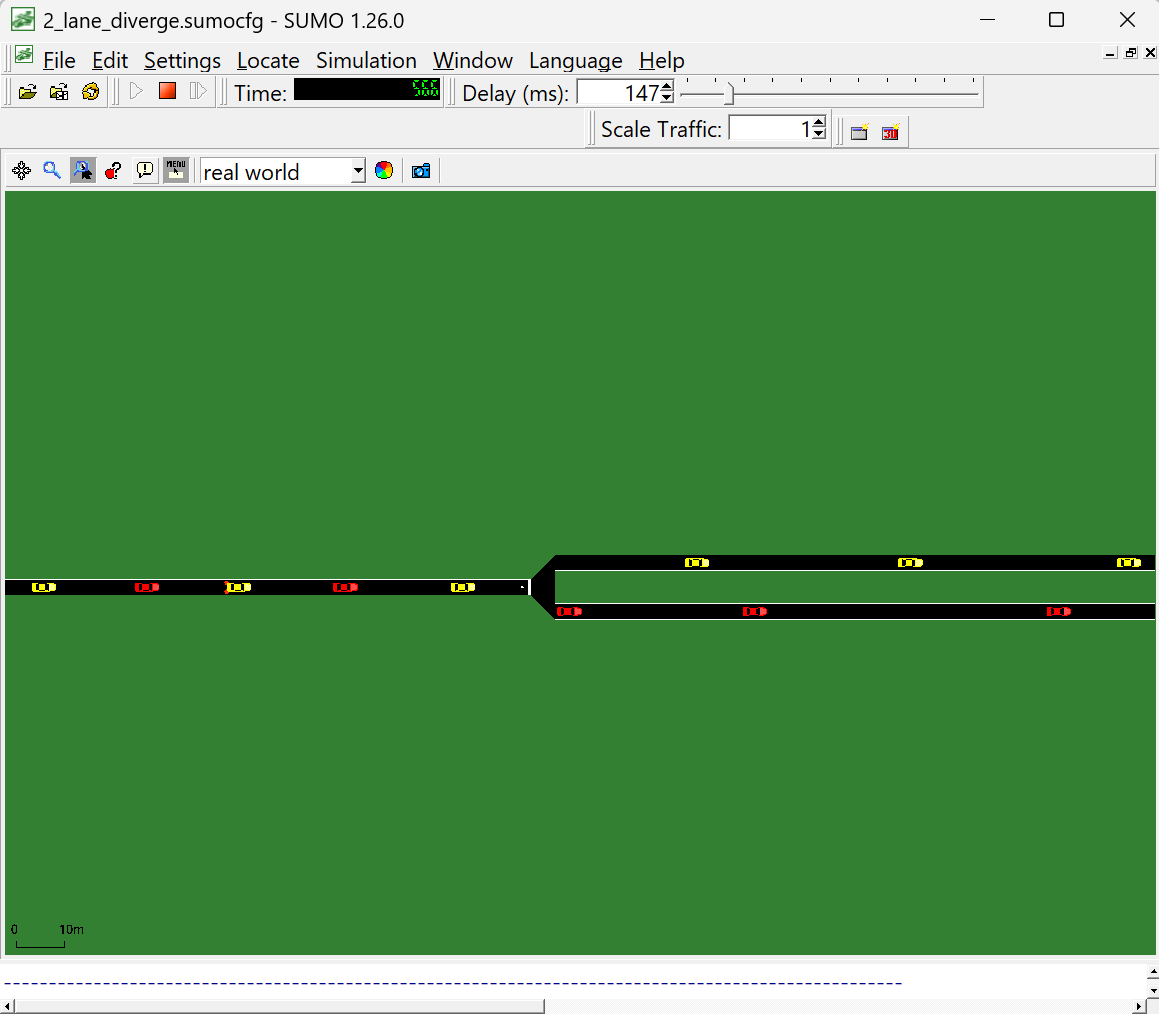}
    \caption{SUMO simulations involving a diverge action.}
    \label{fig:SUMO_diverge}
\end{figure}

 {\bf Validation in SUMO:}
 Fig.~\ref{fig:SUMO_diverge} illustrates the SUMO simulation of the single lane diverge scenario, where vehicles traveling on road $P$ split into roads $A$ and $B$. Here, $\gamma^A=\gamma^B=0.5$. The snapshot shows vehicle interactions and accumulation near the divergence point as the incoming traffic stream splits into two downstream flows.
\begin{figure}[htbp]
    \centering

    \begin{subfigure}[b]{0.49\textwidth}
        \centering
        \includegraphics[width=\textwidth]{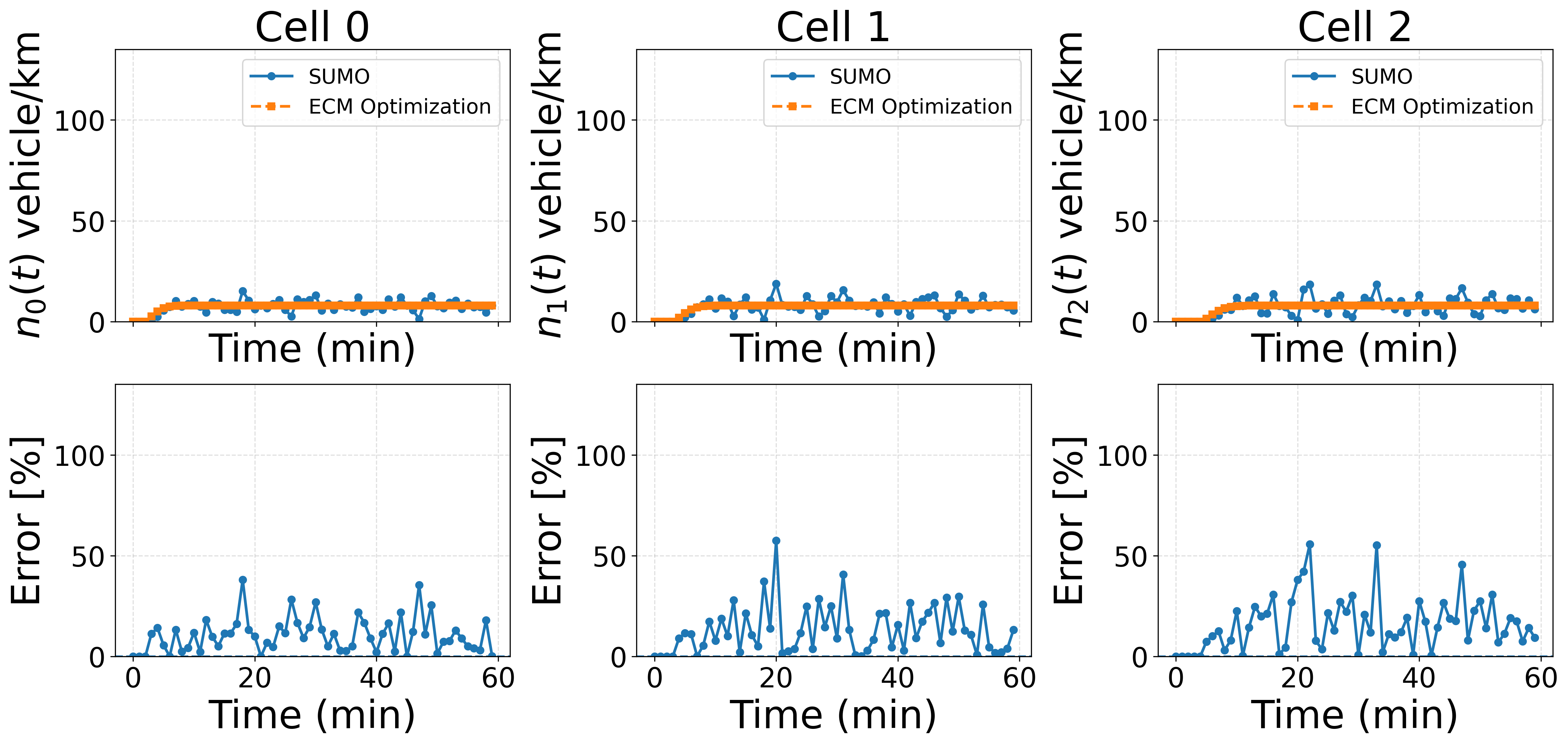}
        \caption{Road $A$}
        \label{fig:divergeA}
    \end{subfigure}
    \hfill
    \begin{subfigure}[b]{0.49\textwidth}
        \centering
        \includegraphics[width=\textwidth]{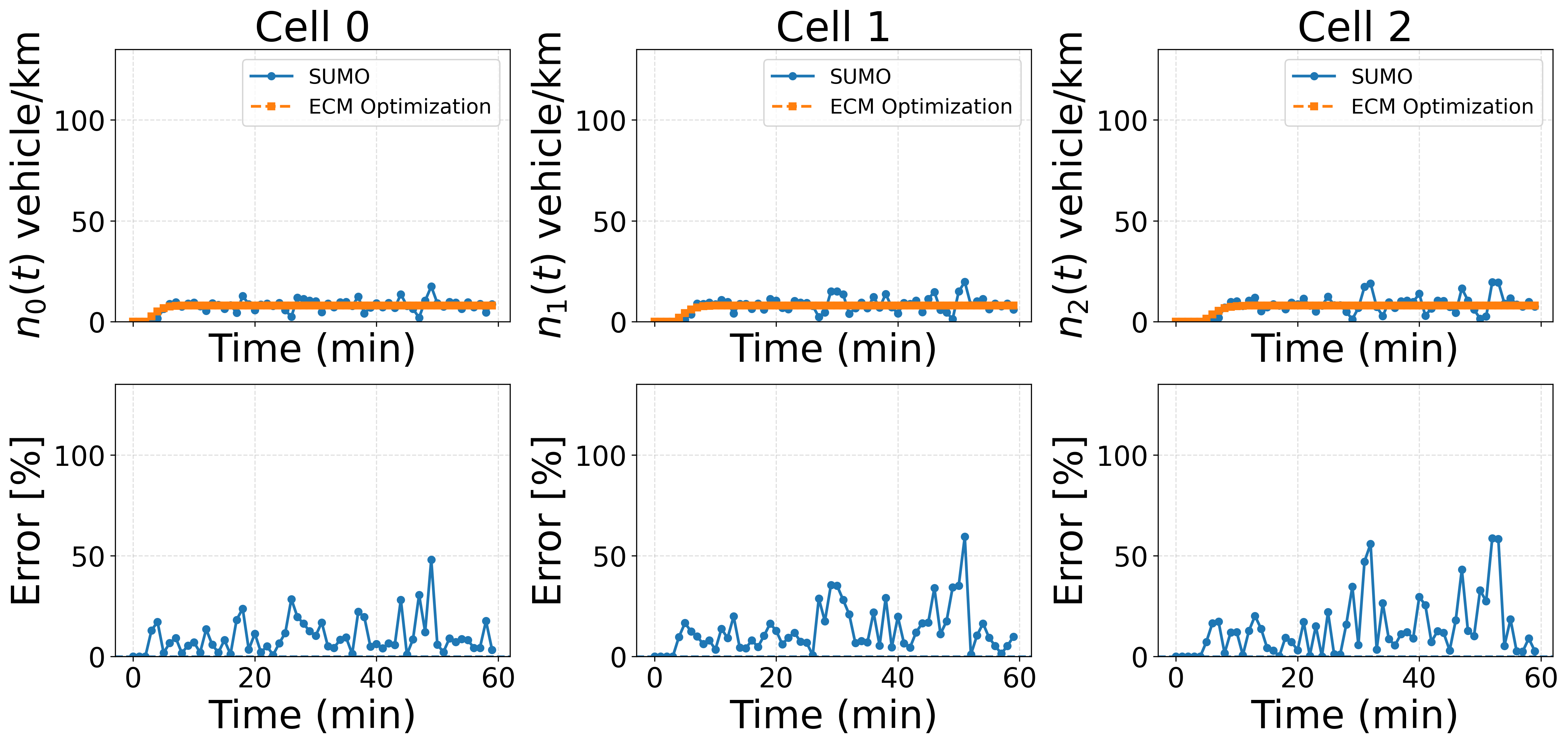}
        \caption{Road $B$}
        \label{fig:divergeB}
    \end{subfigure}
    \hfill
    \begin{subfigure}[b]{0.49\textwidth}
        \centering
        \includegraphics[width=\textwidth]{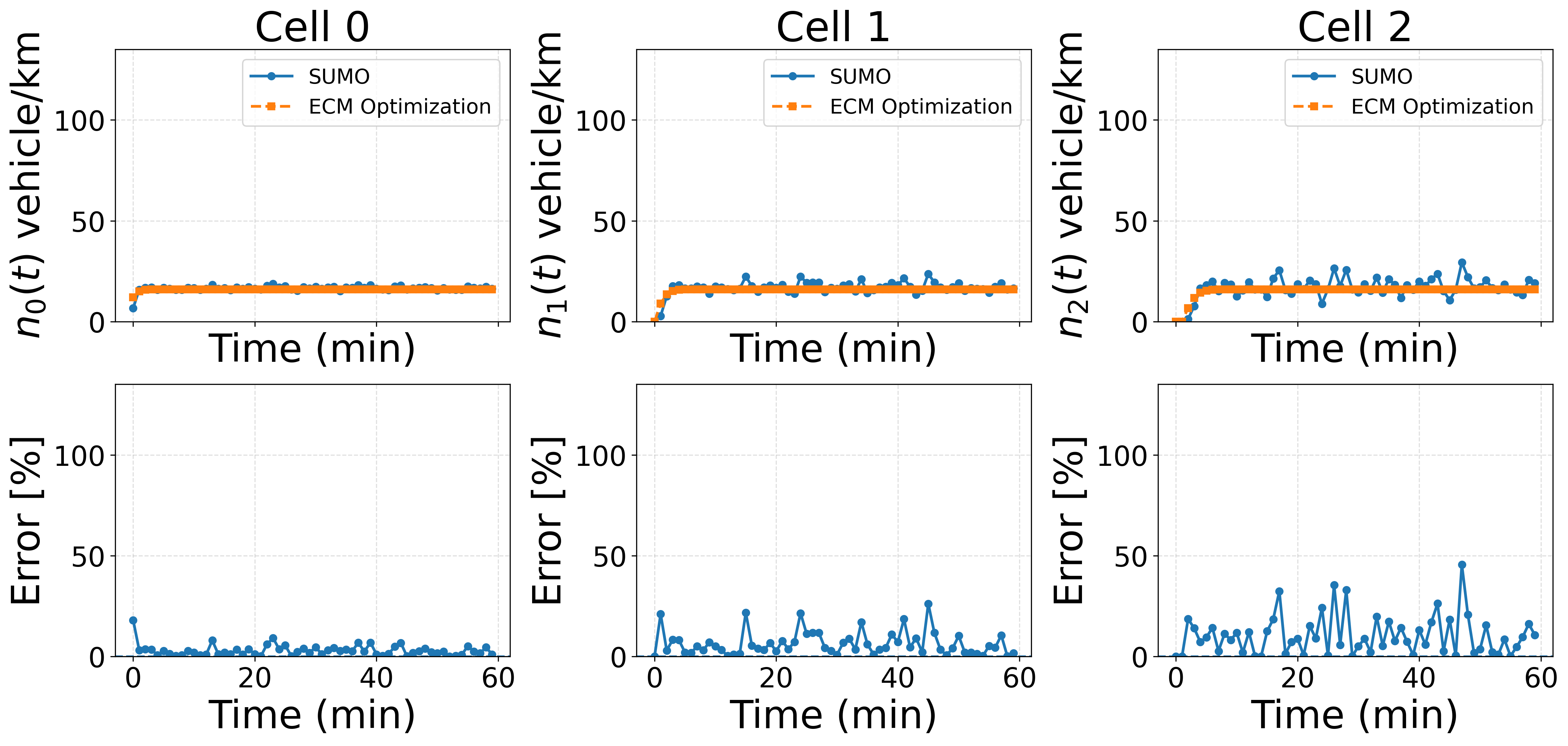}
        \caption{Road $P$}
        \label{fig:divergeP}
    \end{subfigure}

    \caption{Comparison of vehicle concentration values for different road cells obtained from the SUMO simulation and ECM optimization, along with the corresponding concentration errors for the diverging simulation.}
    \label{fig:diverge_results}
\end{figure}

Fig.~\ref{fig:diverge_results} presents the validation results for the single-lane diverge scenario by comparing the vehicle concentrations obtained from the proposed ECM with those generated by the microscopic SUMO simulation. Fig. 20a, 20b, and 20c correspond to roads A, B, and P, respectively. Each figure contains three concentration plots corresponding to Cells 0, 1, and 2, along with their respective percentage-error plots.

Fig.~\ref{fig:divergeA} and \ref{fig:divergeB} compare the SUMO and ECM concentrations in Cells 0, 1, and 2 of the downstream roads A and B, respectively. Both roads exhibit similar concentration patterns due to the equal traffic splitting ratios, $\gamma^A=\gamma^B=0.5$. At the beginning of the simulation, the vehicle concentrations in all three cells increase as vehicles traveling along the upstream road P enter the downstream roads. Unlike the merging scenario discussed in Section~\ref{section:merge}, where two incoming traffic streams compete for the limited receiving capacity of a single downstream road, the diverging scenario distributes the incoming traffic between two outgoing roads. Consequently, the vehicle concentrations in roads A and B remain relatively low throughout the simulation, indicating that both roads have sufficient capacity to accommodate their respective incoming flows. The ECM reproduces the overall concentration trends observed in SUMO, particularly the initial increase and subsequent stabilization of vehicle concentrations. However, SUMO exhibits more frequent short-term fluctuations, whereas the ECM produces smoother concentration trajectories. These differences are reflected in the percentage-error plots, where intermittent error peaks occur throughout the simulation. Such discrepancies can be attributed to the microscopic interactions between individual vehicles in SUMO, including variations in vehicle arrival times, speeds, and intervehicle spacing, which are averaged out in the macroscopic ECM representation.

Fig.~\ref{fig:divergeP} presents the concentration profiles for Cells 0–2 of the upstream road P, where vehicles enter the network before splitting into roads A and B. Unlike the merging scenario, where the incoming roads experience significant vehicle accumulation due to downstream capacity restrictions, road P maintains relatively low vehicle concentrations throughout the simulation. This behavior can be attributed to the ability of the two downstream roads to accommodate the traffic distributed from road P, thereby preventing substantial congestion near the divergence point. The concentration in Cell 0 initially increases before approaching a nearly constant value, indicating steady traffic flow toward the downstream cells. Similar trends are observed in Cells 1 and 2, although SUMO exhibits greater fluctuations, particularly near the divergence point, due to microscopic vehicle interactions. In contrast, the ECM produces smoother concentration profiles while closely capturing the overall trends observed in SUMO. The percentage-error plots show relatively small discrepancies in Cell 0, with larger transient errors in Cells 1 and 2 arising from microscopic fluctuations not explicitly captured by the ECM.

Overall, Fig.~\ref{fig:diverge_results} demonstrates that the proposed ECM successfully captures the principal macroscopic traffic behavior associated with a single-lane diverge, including vehicle accumulation, traffic distribution between downstream roads, and the evolution toward approximately steady concentration levels. Although the ECM does not reproduce the short-term fluctuations observed in SUMO, it provides a reasonable approximation of the overall concentration dynamics across all three roads. These results, together with the merging scenario presented in Fig. 16, demonstrate the capability of the proposed ECM framework to represent traffic interactions at both merging and diverging network configurations.
\section{Conclusion}
This paper presented an ECM for dynamic traffic modeling by exploiting analogies between electrical and transportation variables. The proposed formulation represents vehicle concentration as capacitor voltage and traffic flow as voltage-dependent controllable current sources. We showed that this circuit is equivalent to a space-discretized LWR.  The framework was extended to account for restricted network inflow, traffic signals, and merging and diverging traffic streams. Validation against microscopic SUMO simulations showed that the ECM captures key traffic phenomena, including queue formation, upstream congestion propagation, saturation, and recovery under different demand levels. For the two-lane merge scenario and single-lane diverge scenario, the model also reproduced the overall concentration trends observed in SUMO, while naturally smoothing short-term microscopic fluctuations. These results support the ECM as a compact and physically interpretable framework for dynamic transportation-network modeling and future traffic optimization and control applications.

Future work will focus on extending the proposed ECM framework to more complex transportation-network configurations and improving its quantitative agreement with microscopic traffic simulations. 

\section*{Appendix}

\subsection*{B. Small-signal ECM (ss-ECM)}
In this section, we propose a ``small-signal" ECM which is valid for small perturbations around the steady-state conditions derived in Section~\ref{section:steady-state}. We will call this model ``ss-ECM".

\textbf{Linearization}: Let us introduce deviations with respect to the steady-state conditions:
\begin{eqnarray}
    && \Delta q_{in} = q_{in}-q^*_{in}, \quad
    \Delta n_i = n_i - n_i^*, \quad 
    \Delta q_i = q_i - q_i^* \nonumber \\
    && \Delta q_{out} = q_{out} - q_{out}^*
\end{eqnarray}

The ssECM aims to capture the dynamic behavior of $n_i(t)$ when small input ($\Delta q_{in}$) and output flow perturbations ($\Delta q_{out}$) are introduced in the network. This dynamic behavior can be approximated using a first-order Taylor series expansion around the eq. point
\begin{eqnarray}
    \frac{d}{dt} n_i &\approx& 
    \overbrace{
   \frac{\partial q_i}{\partial n_{i-1}}\bigg|_*  
   \Delta n_{i-1}+ 
   \frac{\partial q_i}{\partial n_{i}}\bigg|_*  
   \Delta n_{i}}^{ \Delta q_i} \\
   &&    
   - 
   \underbrace{
   \left(\frac{\partial q_{i+1}}{\partial n_i}\bigg|_* \Delta n_i 
   +\frac{\partial q_{i+1}}{\partial n_{i+1}}\bigg|_* \Delta n_{i+1} \right)}_{\Delta q_{i+1}}
\end{eqnarray}
for $i\in \{1,\ldots, N-1\}$ 
where the partial derivatives are evaluated at the steady-state conditions. 
A similar approximation can be derived for the boundary cells $i\in \{0, N\}$, which we omit here for brevity. 
Two regimes are considered: i) no congestion; and ii) congestion.

\begin{figure*} 
    \centering
    \includegraphics[width=0.9\linewidth]{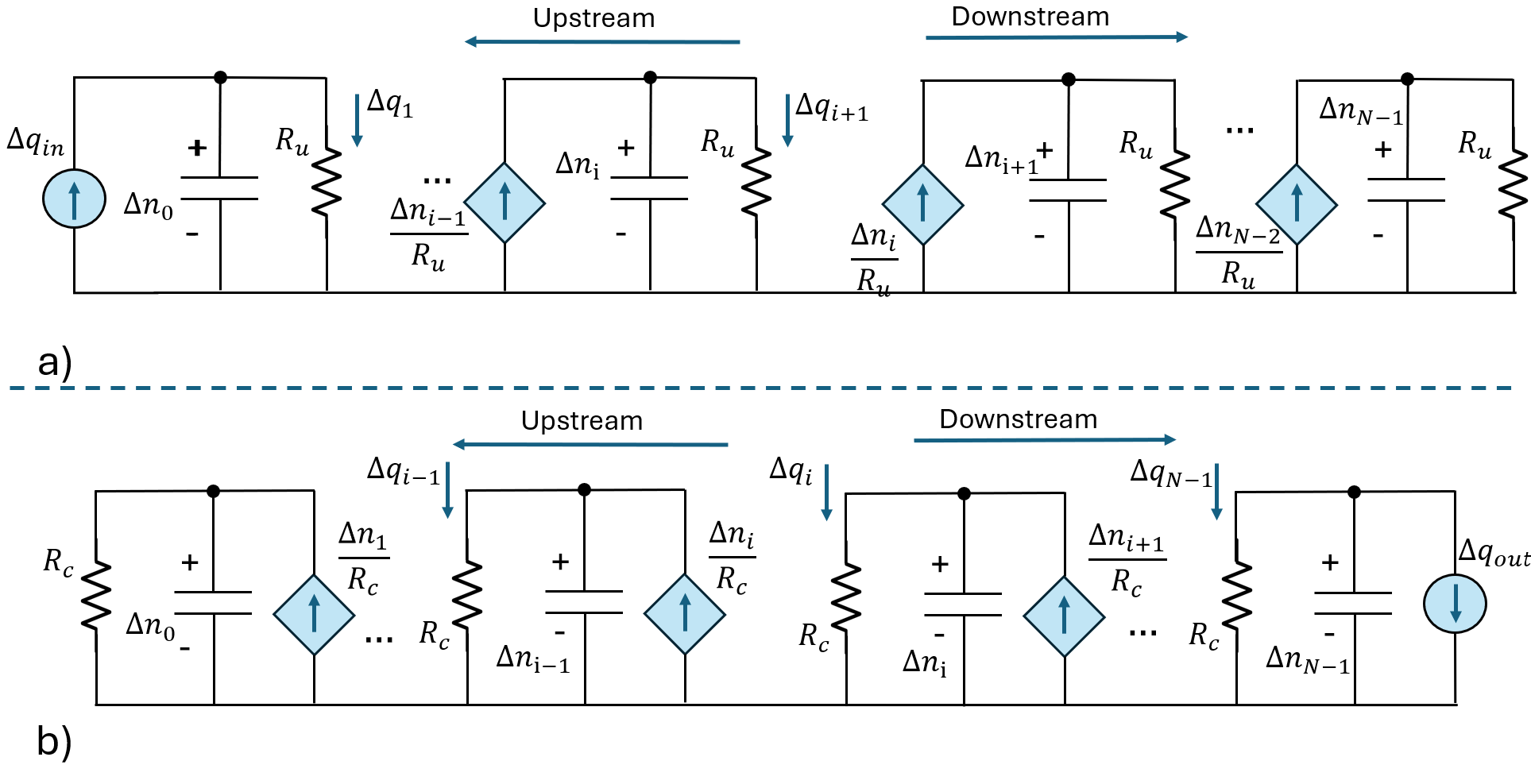}
    \caption{Small signal ECM for non-congested road (a) and congested road (b). }
    \label{fig:smallSignalECM}
\end{figure*}

\textbf{Regime I (No congestion)}:
The dynamical model in this regime can be approximated as:
\begin{eqnarray}
    \frac{d}{dt} n_0
    &\approx& \Delta q_{in} -  \frac{1}{R_u} \Delta n_0 \\
    \frac{d}{dt} n_i &\approx & 
     \frac{1}{R_u} \Delta n_{i-1} -  \frac{1}{R_u} \Delta n_{i} \\
         \frac{d}{dt} n_{N-1} &\approx & 
     \frac{1}{R_u} \Delta n_{N-2} -  \frac{1}{R_u} \Delta n_{N-1} 
\end{eqnarray}
where $R_u=\Delta x / u^{free}$ is denoted the "uncongested" resistance, $\Delta x = 1$ km and $i \in \{ 1, \ldots, N-2\} $. 
These equations can be realized by the ssECM proposed in  Figure~\ref{fig:smallSignalECM}(a), which is composed of a series of RC circuits interconnected by current-controlled sources. 
We note that controllable current sources are dependent on the capacitor voltages ``upstream" of the circuit. Additionally,  the variation in input current $\Delta q_{in}$ (placed at the first cell) is the main excitation in the circuit. As a result, changes in the input flow are propagated ``downstream" during non-congested regimes. It is interesting to note that the resistance $R_u$ depends on the length of the cell and traffic speed. Faster free-flow speeds ($ u^{free}$) lead to a reduction in the resistance and quicker responses/propagation. 

\textbf{Regime II (Congested):}
The dynamical model in this regime can be approximated as:
\begin{eqnarray}
    \frac{d}{dt} n_0 
    &\approx& 
    -\frac{1}{R_c} \Delta n_0 + \frac{1}{R_c} \Delta n_1      \\
    \frac{d}{dt}  n_i &\approx & 
     -\frac{1}{R_c} \Delta n_i + \frac{1}{R_c} \Delta n_{i+1} \\
    \frac{d}{dt} n_{N-1} &\approx & 
     -\frac{1}{R_c} \Delta n_{N-1} - \Delta q_{out}     
\end{eqnarray}
where $R_c=\Delta x / w$ is the congested resistance, $\Delta x = 1$ km and $i \in \{ 1, \ldots, N-2\} $. 
These equations can be realized by the ssECM proposed in  Figure~\ref{fig:smallSignalECM}(b). This circuit is also composed of RC circuits connected with voltage-controlled current sources. However, the location of the resistors and current sources is swapped. 
We note that controllable current sources are dependent on the capacitor voltages "downstream" of the circuit. Additionally,  the variation in our current $\Delta q_{out}$ (placed at the last cell) is the main excitation in the circuit. As a result, changes in the output flow are propagated "upstream" (i.e., backwards in the space) in the circuit during congested regimes.

\bibliographystyle{IEEEtran}
\bibliography{reference}
\end{document}